\documentclass[9pt,article,twoside]{rmaa-rho-class/rmaa-rho}
\RMxAAtemplatetype{\RMxAA} 

\vol{61}
\pages{30-36}
\thisyear{2026}
\doi{\href{https://www.astroscu.unam.mx/rmaa/RMxAA..XX-X}{https://www.astroscu.unam.mx/rmaa/RMxAA..XX-X}}

\usepackage{longtable}

\title{Spectral classification of brown dwarfs using machine learning}

\author[1,3$\dagger$]{A. R. Callen \orcidlink{0000-0001-8600-4798}}
\author[2,$\dagger$]{I. Bustos Fierro \orcidlink{0000-0002-2653-1120}}
\author[2,3]{M. Gómez \orcidlink{0000-0001-6208-9109}}

\affil[1]{Facultad de Matemática, Astronomía, Física y Computación (FAMAF). Av. Medina Allende s/n, Ciudad Universitaria. Córdoba, 5000, Córdoba, Argentina}
\affil[2]{Observatorio Astronómico, Universidad Nacional de Córdoba. Laprida 854, Córdoba, 5000, Córdoba, Argentina }
\affil[3]{Consejo Nacional de Investigaciones
Científicas y Técnicas (CONICET). Godoy Cruz 2290, Buenos Aires, 1425, CABA, Buenos Aires, Argentina.}

\leadauthor{Callen et al.}
\smalltitle{LaTex Macro RMxAA}

\corres{Iv\'an Bustos Fierro}
\email{ivan.bustos.fierro@unc.edu.ar }

\received{April 16th, 2024}
\accepted{\today}

\license{Texto de la licencia aquí}

\setbool{rho-abstract}{true} 
\setbool{rho-resumen}{true} 

\begin{abstract}
Brown dwarfs are compact objects that do not reach temperatures high enough to produce sustained hydrogen fusion. Consequently, they cool over time, gradually evolving through later spectral types. In fact, three new spectral types (L, T, and Y) were added to the Harvard sequence to accommodate the spectral features of brown dwarfs. During the cooling process, some brown dwarfs unexpectedly become bluer instead of redder (at optical and near-infrared wavelengths). This phenomenon, known as the bluing effect, is particularly noticeable at the L/T spectral transition. The aim of this work is to approximate the spectral type of brown dwarfs using only photometric data, in particular 2MASS and WISE magnitudes. We used two machine learning algorithms, Random Forest and Gaussian Processes, which were evaluated using a 70/30 train/test split. Both models were trained using 5-fold cross-validation and achieved F1-scores of 0.84 and 0.87, respectively, on the test set. After validating the reliability of the algorithms, we applied them to 21 isolated brown dwarfs without prior spectral type determinations. Our results indicate that 5 of these objects have a spectral type between L0 and L4, while the remaining 16 fall within the M6-M9 range. Machine learning algorithms, combined with multi-band photometry, are a powerful tool for estimating the spectral types of brown dwarfs.
\end{abstract}


\begin{resumen}
Las enanas marrones son objetos compactos que no alcanzan temperaturas suficientemente altas para producir una fusión de hidrógeno sostenida. En consecuencia, se enfrían con el tiempo, evolucionando gradualmente hacia tipos espectrales más tardíos. De hecho, se a\~nadieron tres nuevos tipos espectrales (L, T y Y) a la secuencia de Harvard para acomodar las características espectrales de las enanas marrones. Durante el proceso de enfriamiento, algunas enanas marrones se vuelven inesperadamente más azules en lugar de más rojas (en longitudes de onda ópticas e infrarrojas cercanas). Este fenómeno, conocido como el efecto de azulamiento, es particularmente notable en la transición espectral L/T. El objetivo de este trabajo es aproximar el tipo espectral de las enanas marrones utilizando únicamente datos fotométricos, en particular magnitudes de 2MASS y WISE. Empleamos dos algoritmos de aprendizaje automático,  Random Forest y Gaussian Processes, que fueron evaluados utilizando una sub-muestra de entrenamiento/prueba de 70/30. Ambos modelos fueron entrenados utilizando una validación cruzada con 5 particiones y lograron puntuaciones F1 de 0,84 y 0,87, respectivamente, en el conjunto de prueba. Tras validar la fiabilidad de los algoritmos, los aplicamos a 21 enanas marrones aisladas sin determinaciones previas de tipo espectral. Nuestros resultados indican que 5 de estos objetos tienen un tipo espectral entre L0 y L4, mientras que los 16 restantes caen dentro del rango M6-M9. Los algoritmos de aprendizaje automático, combinados con la fotometría multi-banda, son una herramienta poderosa para estimar los tipos espectrales de las enanas marrones.
\end{resumen}

\begin{document}
\nolinenumbers
\maketitle
\pagestyle{fancy}\thispagestyle{firststyle}


\section{INTRODUCTION}
Brown dwarfs are substellar objects with masses intermediate between those of stars and giant planets. They were first conceptualized in 1963 under the name black dwarfs, when \citet{1963ApJ...137.1121K} numerically showed that, for a given chemical composition, there exists a mass limit below which a collapsing object cannot ignite sustained hydrogen fusion in its core. These substellar objects become supported by electron degeneracy pressure. The term brown dwarf (BD) was later introduced in 1975 to specifically refer to such intermediate-mass objects, whereas black dwarf was reserved for stars at the end of their evolution \citep{1975PhDT.........1T}.

Regarding their typical characteristics, brown dwarfs have masses,  approximately,  between 13 and 80 Jupiter masses ($M_{\rm Jup}$). The lower limit corresponds to the onset of deuterium burning, while the upper limit marks the minimum mass required for hydrogen fusion. However, these limits are not universal: the minimum mass, effective temperature, and luminosity required for sustained hydrogen fusion vary with the metallicity of the object, increasing for lower metallicities \citep{2014AJ....147...94D}. For instance, at solar metallicity, the hydrogen-burning limit lies around $0.07-0.074M_{\odot}$ $\equiv$ $73$–$78M_{\rm Jup}$, whereas at zero metallicity it increases to $\sim0.092M_{\odot}$ $\equiv$ $96M_{\rm Jup}$ \citep{RevModPhys.73.719}.

\citet{1995Natur.378..463N} defined brown dwarfs as compact objects that fail to sustain hydrogen fusion. Nevertheless, the most massive brown dwarfs ($\sim60$–$70 M_{\rm Jup}$) may undergo limited hydrogen burning \citep{1993RvMP...65..301B}. Due to their insufficient mass, these reactions eventually cease, and brown dwarfs cool continuously throughout their lives \citep{2020A&A...637A..38P}. As a result, they are intrinsically faint ($L \leq 10^{-3}L_{\odot}$; \citealp{2015ApJ...810..158F}) and cool ($T_{\rm eff} \leq 2700 \text{K}$; \citealp{2021ApJS..253....7K}).

Given their low temperatures, brown dwarfs emit radiation primarily at infrared wavelengths and are thus best studied with infrared surveys such as 2MASS \citep[The Two Micron All-Sky Survey;][]{2006AJ....131.1163S} and WISE \citep[The Wide-Field Infrared Survey Explorer;][]{2010AJ....140.1868W}. Their spectral classification extends beyond the M-type stars and includes three new types: L, T, and Y. In the near-infrared spectral region,  L-type dwarfs exhibit absorption bands of water vapour $(H_2O)$. T-type dwarfs are dominated by strong methane $(CH_4)$ absorption bands, and Y-type dwarfs, the coldest known, show ammonia $(NH_3)$ features, with effective temperatures possibly as low as $\sim250$ K. \citep{1997Sci...276.1350K,2005ARA&A..43..195K,2024MNRAS.533.3784B,2024ApJ...973..107B}.

Unlike main-sequence stars—which exhibit a direct correlation between spectral type and colour indices, appearing progressively redder in diagrams such as $V-R$ vs. $R-I$ \citep{1990A&AS...83..357B, 2012PASP..124..140B}—brown dwarfs display far more complex behaviour. 
Developing a colour-spectral type relation for these substellar objects is highly desirable, as it provides a useful –albeit complex— temperature proxy and a diagnostic tool for atmospheric physics. A notable departure from stellar trends occurs at the L/T transition: the "bluing effect" \citep{2002ApJ...564..452L,2020MNRAS.499..505D,2023AJ....166..103S}, where cooler objects appear paradoxically bluer in specific indices like $J-H$ and $J-K$.

This phenomenon is driven by the evolution of atmospheric chemistry as the object's temperature drops. Refractory elements—such as titanium, vanadium, iron, and silicates—condense into dust and liquid "clouds," which gravity then pulls into deeper, hotter layers. This sedimentation process removes these heavy particles from the upper atmosphere, preventing them from absorbing light \citep{2020A&A...637A..38P, 2021ApJ...920...85M, 2022MNRAS.513.5701S}.  While carbon remains trapped in carbon monoxide ($CO$) in warmer L-dwarfs, this shift in chemical equilibrium allows methane to dominate the spectrum as the transition to T-dwarfs occurs. Consequently, deep methane ($CH_4$) absorption bands, particularly in the $H$ and $K$ bands of the 2MASS filters, suppress flux in those regions and deviate from expected photometric trends \citep{2011ApJS..197...19K}.

Ultimately, although cooler brown dwarfs are not intrinsically bluer, their observed colours are heavily modulated by methane absorption bands. A well-defined calibration of these properties enables the efficient classification of candidates in large-scale photometric surveys where spectroscopic follow-up is unfeasible. Thus, while the colour-spectral type relation is more difficult to interpret than in the main sequence, it remains a far more scientifically rewarding metric for understanding substellar evolution.

This study investigates the relationship between spectral type and colour index in brown dwarfs to determine if their spectral type can be classified using only photometric data, which is valuable due to their intrinsic faintness and the difficulty in obtaining high signal-to-noise spectra. Previous works have reported the complex relation between spectral types and colour indices \citep{2015ApJ...810..158F, 2021ApJS..253....7K} and have used individual polynomial fits of colour indices to assign photometric spectral types to brown dwarfs \citep{2015A&A...574A..78S, 2021ApJS..253....7K}.

Other studies, such as those by \citet{2024AJ....168..211B} and \citet{2025A&A...695A.195K}, prioritize the systematic discovery of new candidates. By leveraging machine learning tools to analyze time-series data and high proper motion images from the WISE survey, these works have streamlined the identification process.

The current work extends this by applying machine learning techniques to multiple colour indices simultaneously, using data from 2MASS and WISE. While such techniques have been used to search for new brown dwarfs \citep{2023A&C....4500744A, 2024arXiv240904490K, 2024AJ....168..211B}, here, we train two supervised learning classifiers —Random Forest and Gaussian Processes— on brown dwarfs with known spectral types to classify those without a prior spectral determination. For different classification tasks Random Forests have been employed by \citet{2011MNRAS.414.2602D, 2018MNRAS.476.3974P, 2022A&A...657A..62K, 2023A&A...679A.127G, 2025ApJ...986...19W}, and Gaussian Processes by \citet{2019AJ....158..257B,2023MNRAS.521.1601B}.

In this context, Section 2 presents the data and methodology used in this study, Section 3 discusses the results and compares them with previous studies, and Section 4 provides a general discussion and summarises the main conclusions of this work.

\section{Data and analysis}

\subsection{The dataset}

For this work, we used the database The UltracoolSheet: Photometry, Astrometry, Spectroscopy, and Multiplicity for 4000+ Ultracool Dwarfs and Imaged Exoplanets\footnote{https://zenodo.org/records/10573247}. This catalogue contains over 4000 ultracool dwarfs (spectral types M6 and later) and includes photometric data from multiple surveys, astrometry, proper motions, parallaxes, spectral types, ages, etc. UltracoolSheet began as a catalogue of all spectroscopically confirmed objects of spectral type L0 or later in the solar neighbourhood (up to 100 pc) known as of April 15, 2015. It was subsequently expanded to include late-M-type dwarfs and has also been updated with more recent discoveries, including all ultracool spectral types. 

\begin{figure}[!ht]
    \centering
    \includegraphics[width=\columnwidth]{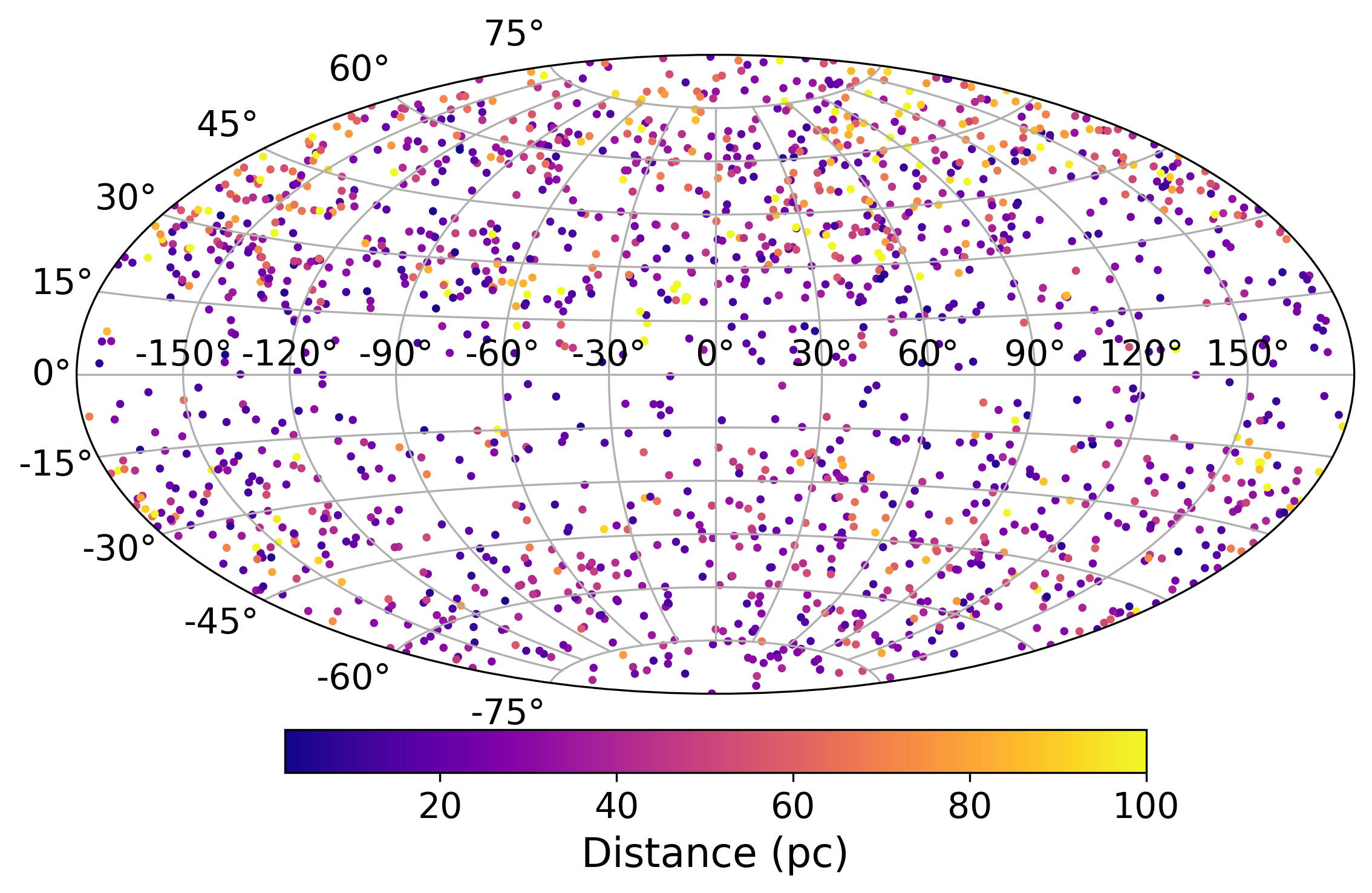}
    \caption{Distribution of the 1723 isolated brown dwarfs in Galactic coordinates. The colour scale indicates the distance to each object in parsecs.}
    \label{proyeccion}
\end{figure}

Specifically, we utilized the version published in February 2024;  however, a newer update was released after the completion of this study, on December 14, 2024. We filtered this catalogue to include only isolated, single objects with masses between 13 and 80 Jupiter masses\footnote{Masses were obtained from the UltracoolSheet catalog, where they are derived using two different evolutionary models: the \citet{2008ApJ...689.1327S} tracks for faint objects ($\log(L/L_{\odot}) \leq -3.75$) and those of \citet{2015A&A...577A..42B} for brighter objects ($\log(L/L_{\odot}) > -3.75$).} and known spectral types. Brown dwarfs identified as having massive companions, as well as those flagged as potential unresolved binaries, were excluded from the sample. Thus, the final sample we used consisted of 1723 isolated brown dwarfs. The distribution of these objects in Galactic coordinates (latitude and longitude) is shown in Figure \ref{proyeccion}, indicating that they are well spread across the sky and located in the solar neighbourhood.

Among the photometric data included in the catalogue, two of the most important sources are the 2MASS and WISE surveys, which provide near- and mid-infrared observations, respectively. 2MASS\footnote{https://irsa.ipac.caltech.edu/Missions/2mass.html} was a survey conducted between 1997 and 2001 that scanned approximately $70\%$ of the sky and detected more than 450 million infrared emitting sources. In order to cover the entire sky, twin 1.3 m diameter telescopes were used, one in the Northern Hemisphere at Mount Hopkins, Arizona, and the other in the Southern Hemisphere at Cerro Tololo, Chile. Each telescope was equipped with detectors capable of simultaneously observing the sky in the J (1.25 $\mu$m), H (1.65 $\mu$m), and Ks (2.17 $\mu$m) bands. 

WISE\footnote{https://wise2.ipac.caltech.edu/docs/release/allsky/} was a NASA space mission launched in 2009. It was tasked with repeatedly mapping the entire sky in infrared wavelengths with a $40 cm$ diameter telescope. One of the key advantages of WISE was its position in space, which allowed it to observe in the mid-infrared, a range inaccessible to ground-based telescopes because of atmospheric absorption. WISE used filters centred at wavelengths of 3.4, 4.6, 12, and 22 $\mu$m (W1, W2, W3, W4), with W1 and W2 specifically designed to probe the 3.3 $\mu$m methane ($CH_4$) absorption band, which is characteristic of brown dwarfs. 

\subsection{Colour-colour and colour-magnitude diagrams}
\begin{figure*}[!ht]
    \centering
    \begin{subfigure}{0.32\linewidth}
        \includegraphics[width=\linewidth]{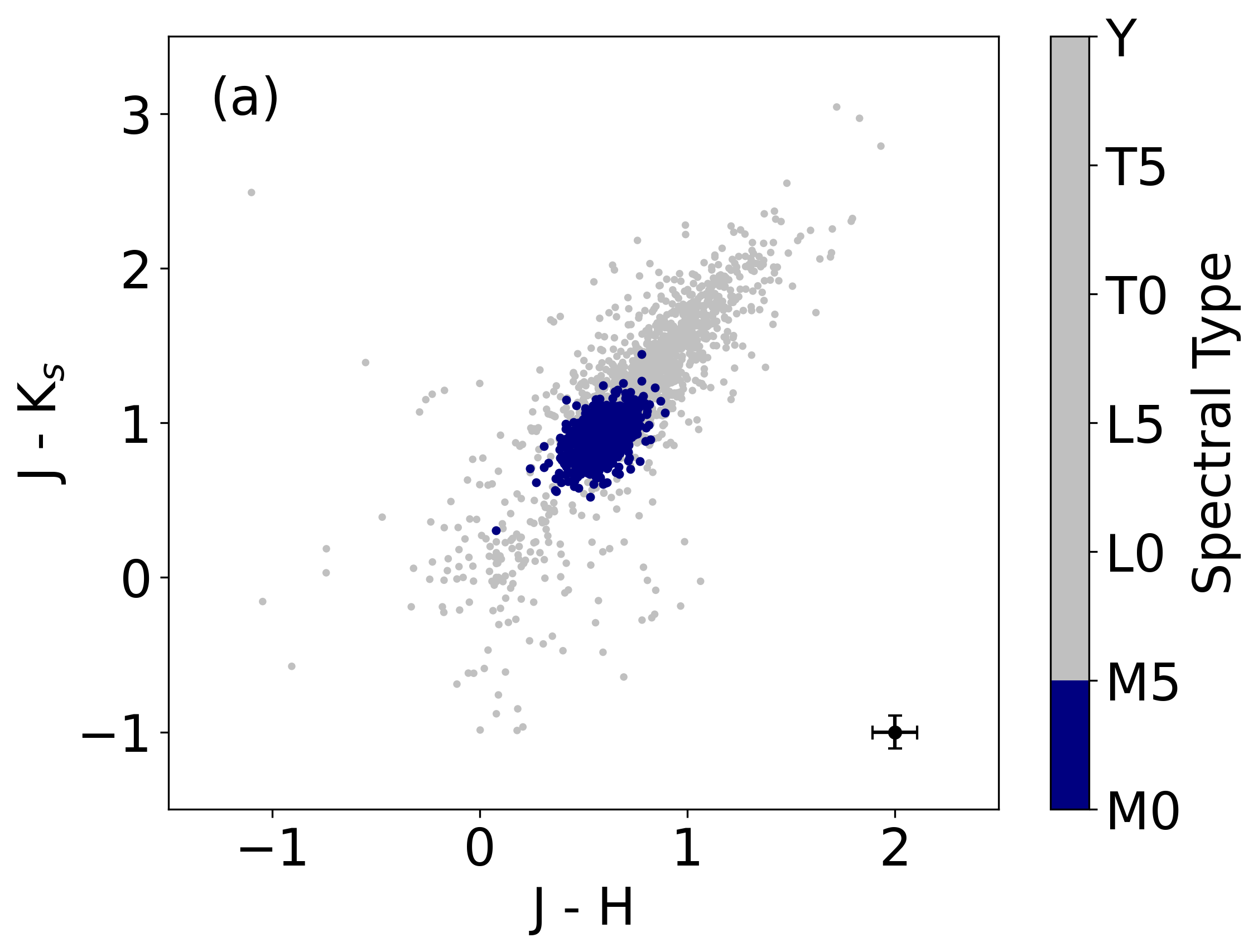}
    \end{subfigure} \hfill
    \begin{subfigure}{0.32\linewidth}\includegraphics[width=\linewidth]{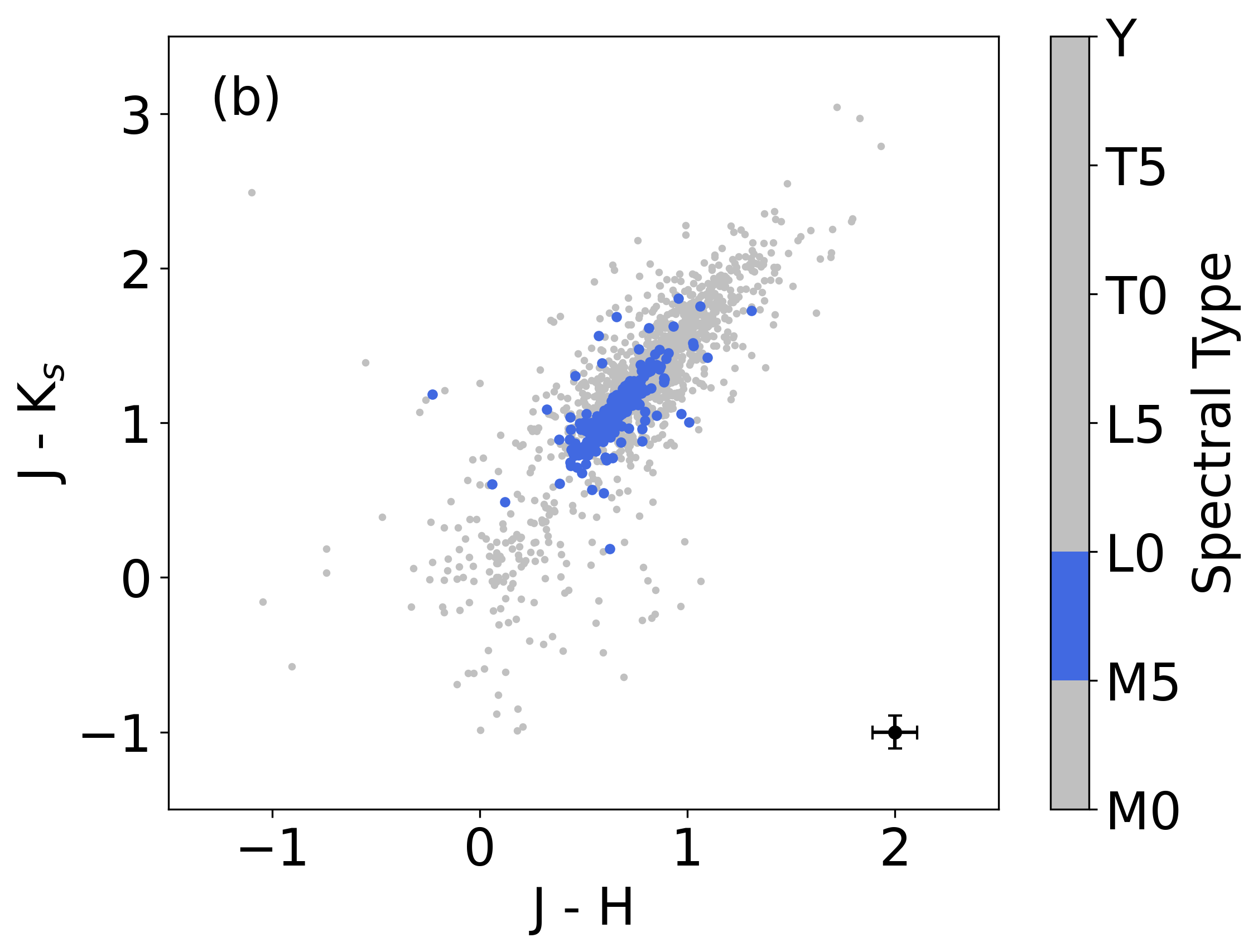}
    \end{subfigure}\hfill
    \begin{subfigure}{0.32\linewidth}\includegraphics[width=\linewidth]{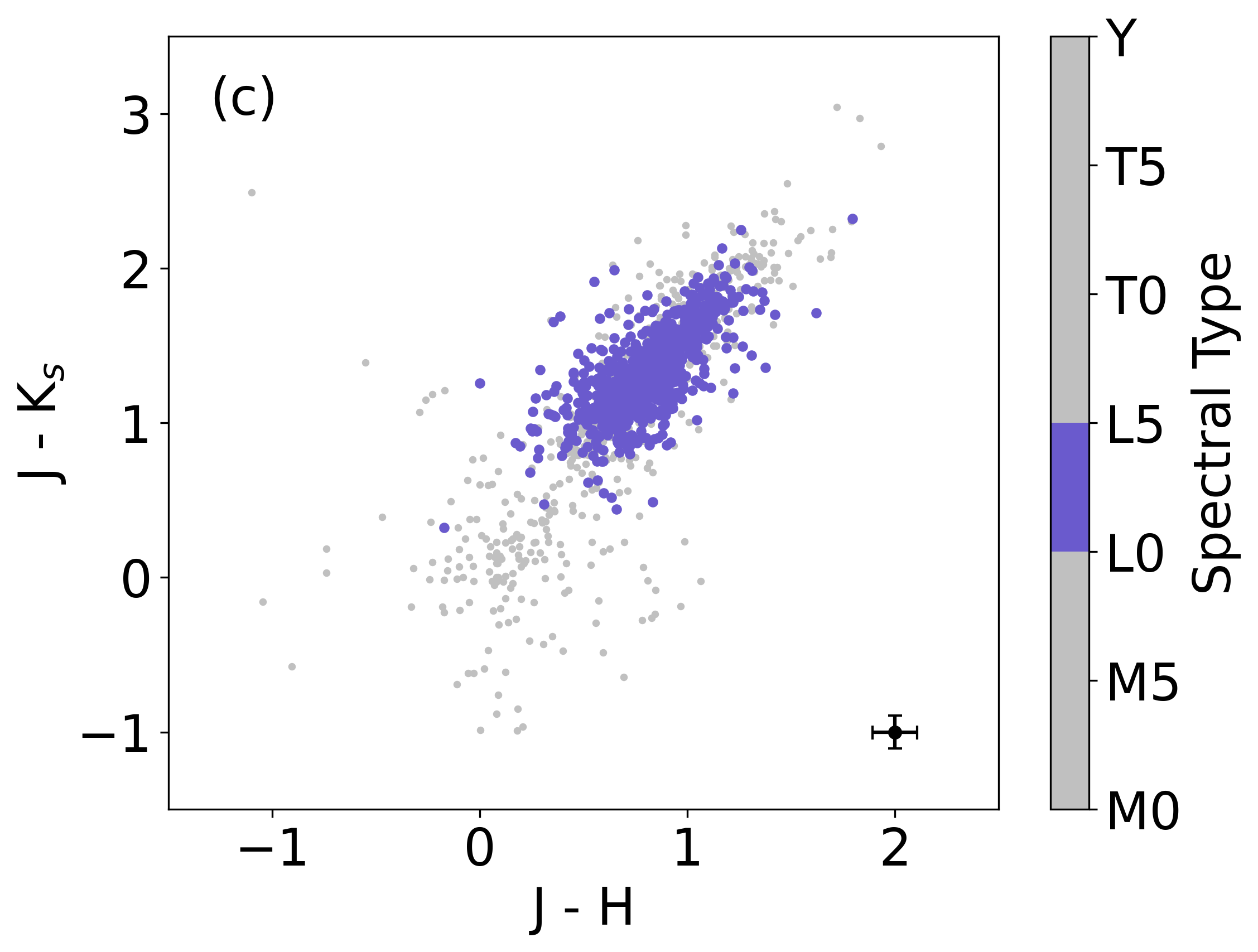}
    \end{subfigure}
    \begin{subfigure}{0.32\linewidth}
    \includegraphics[width=\linewidth]{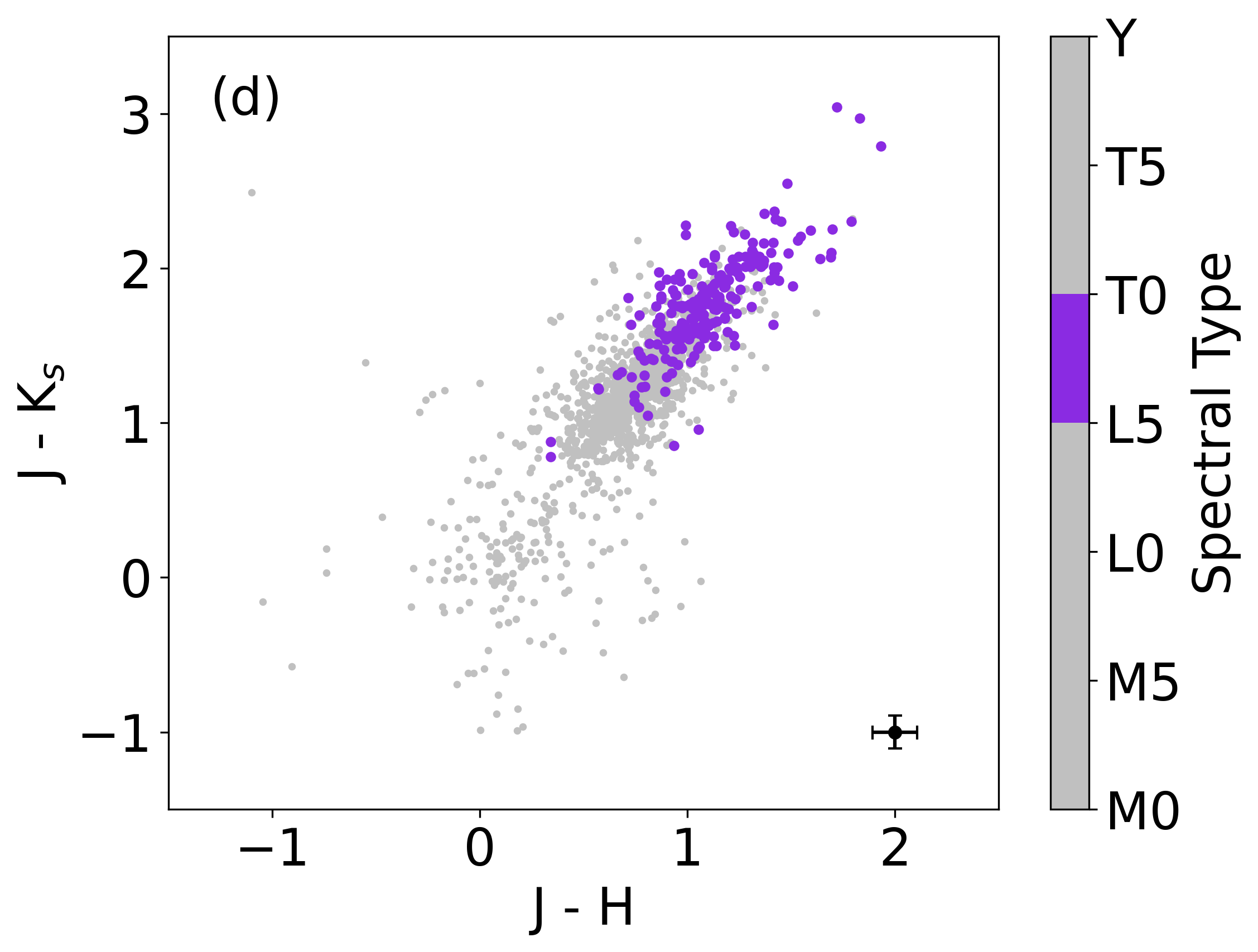}
    \end{subfigure} \hfill
   \begin{subfigure}{0.32\linewidth}
   \includegraphics[width=\linewidth]{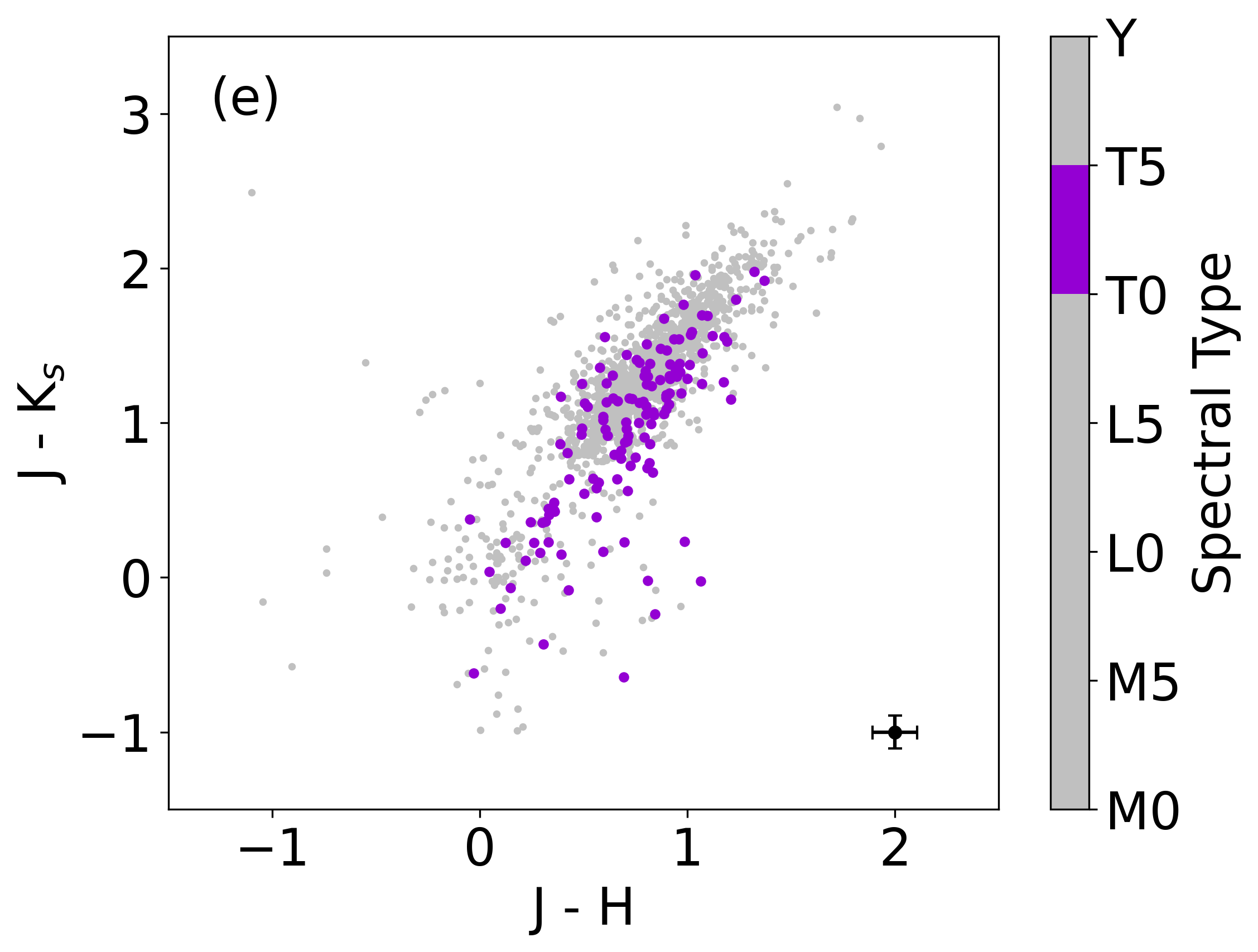}
   \end{subfigure} \hfill
  \begin{subfigure}{0.32\linewidth}
  \includegraphics[width=\linewidth]{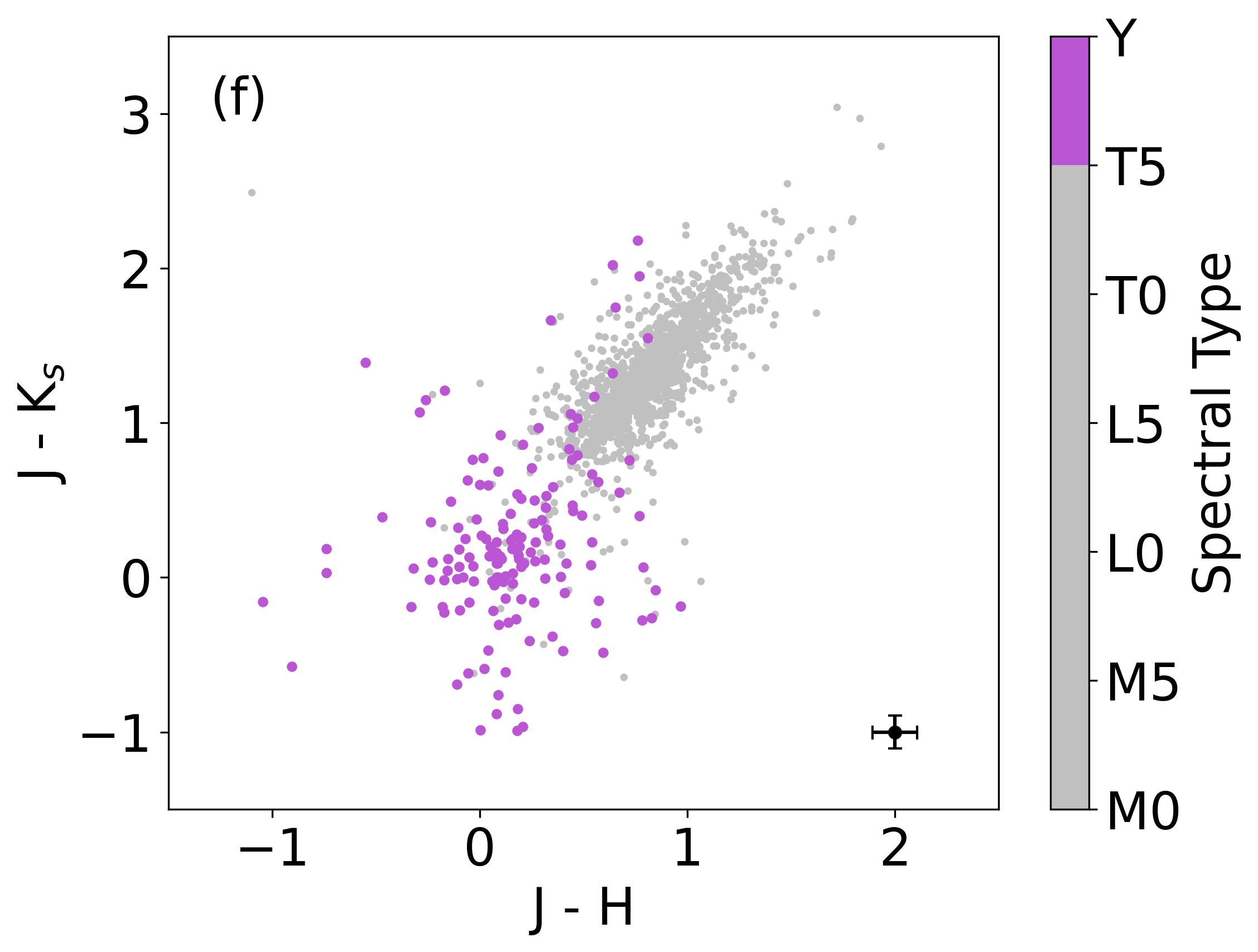}
  \end{subfigure}
    \begin{subfigure}{0.32\linewidth}
    \includegraphics[width=\linewidth]{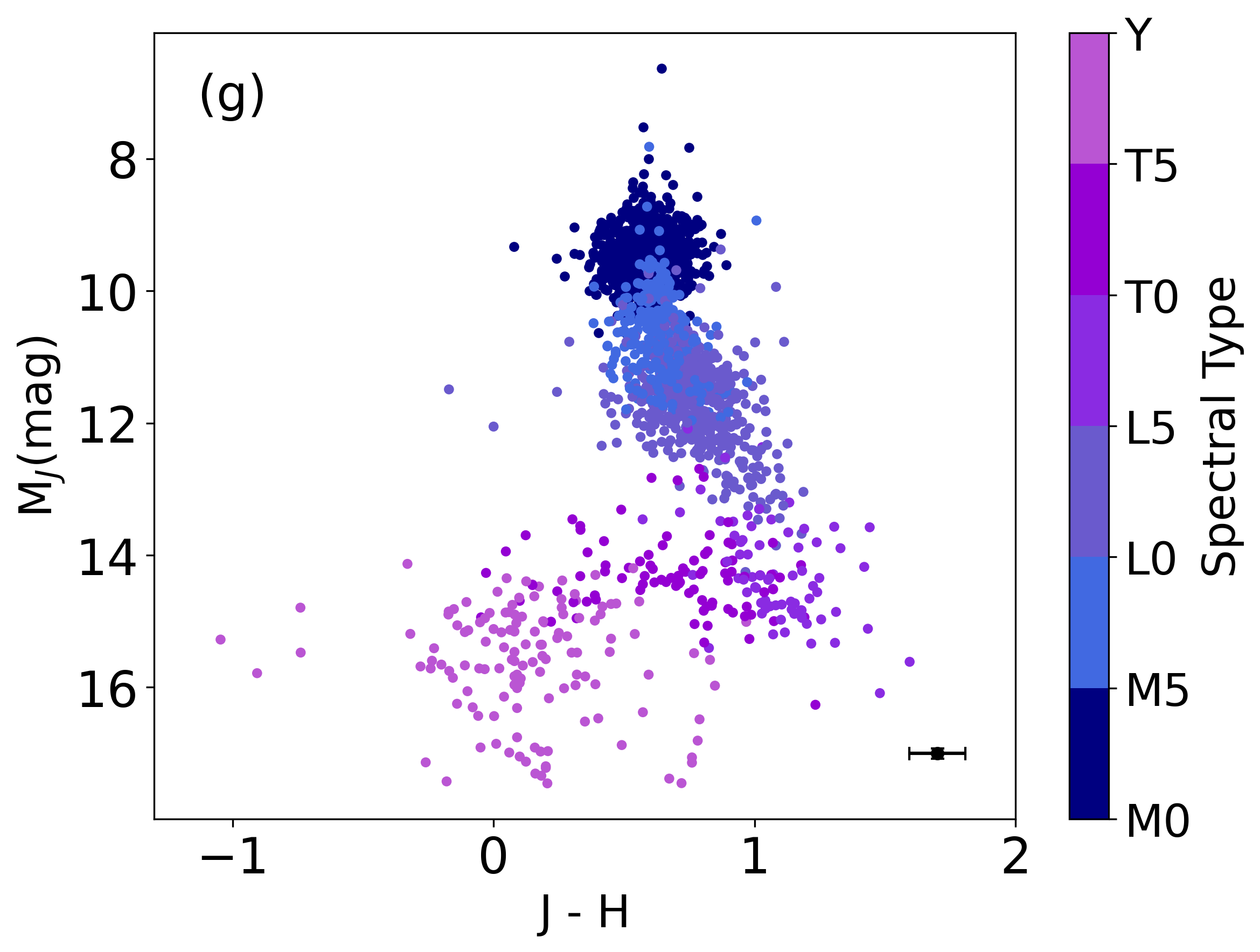}
    \end{subfigure} \hfill
   \begin{subfigure}{0.32\linewidth}
   \includegraphics[width=\linewidth]{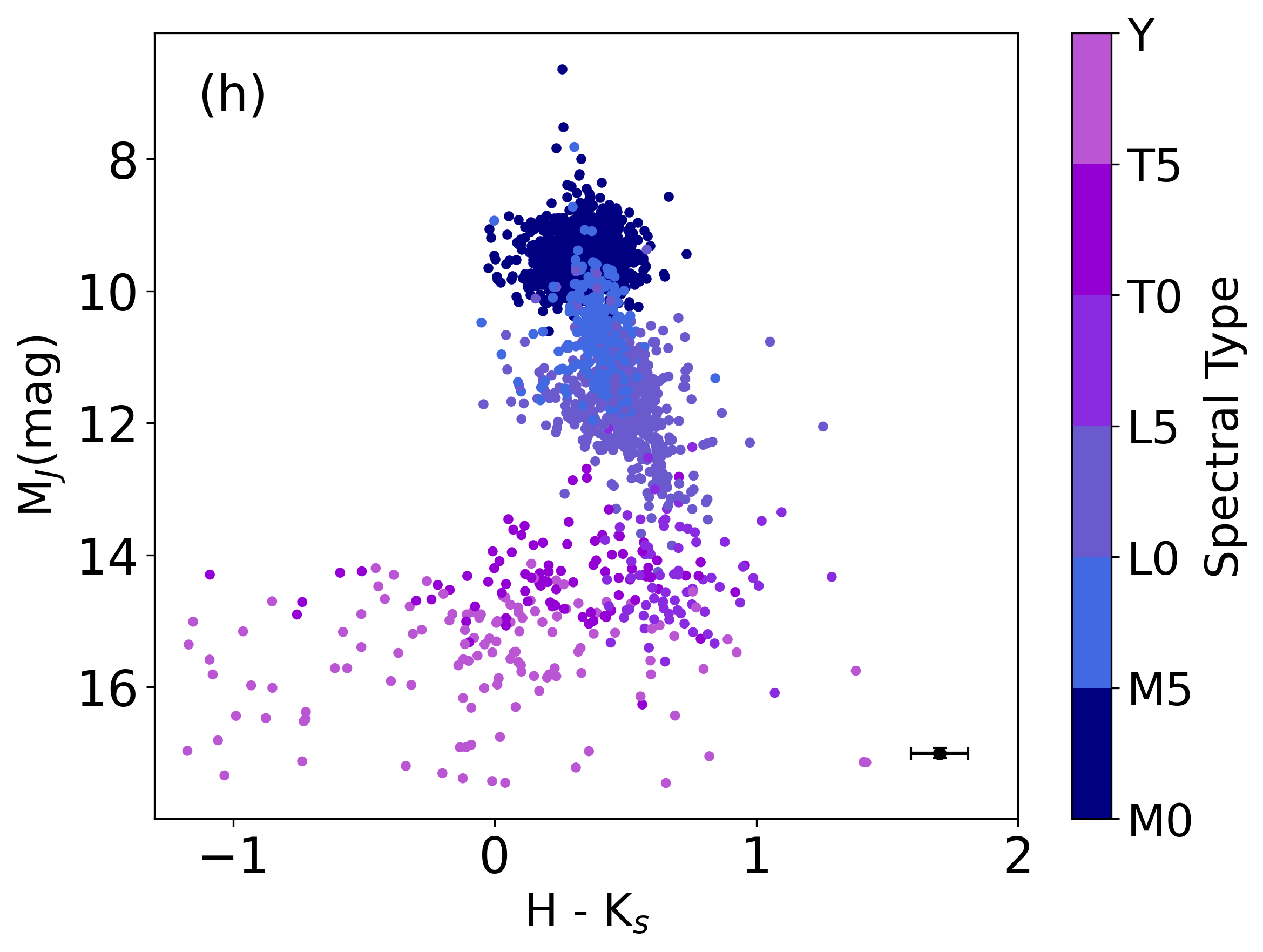}
   \end{subfigure} \hfill
  \begin{subfigure}{0.32\linewidth}
  \includegraphics[width=\linewidth]{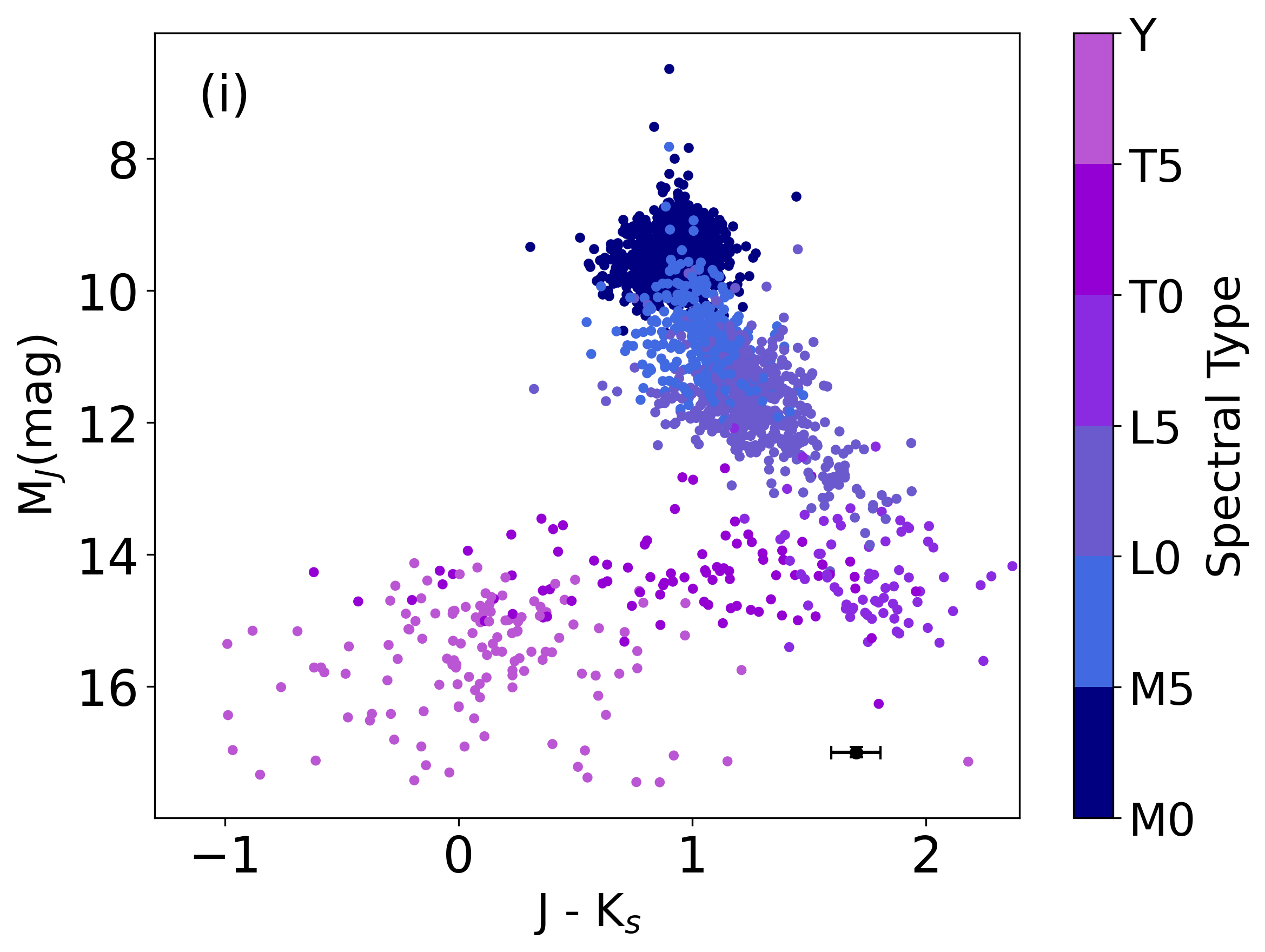}
  \end{subfigure}
    \caption{Top and middle panels: Colour–colour diagrams (J$-$K vs. J$-$H) as a function of spectral type. The spectral types highlighted in each panel are as follows: M0-M4 (a), M5-M9 (b), L0-L4 (c), L5-L9 (d), T0-T4 (e), T5-Y (f). The remaining objects in the sample are shown in grey for reference. Bottom panels: colour–magnitude diagrams, $M_J$ versus $(J-H)$, $(H-K_s)$, and $(J-K_s)$, corresponding to panels (g), (h), and (i). The black point with error bars in the lower-right corner of each panel represents the median uncertainty for all objects in the sample. Note: Types M0–M5 are shown for contextual continuity but are not part of the spectral classes defined in this work. The T5–Y group in panel (f) is further subdivided into Class 4 (T5–T9) and Class 5 (Y) in our classification scheme (see Table 1).}
    \label{jhk-por-te}
\end{figure*}

Using photometric data from the surveys described in the previous section, we explored the relationship between different colour indices and the spectral type\footnote{The UltracoolSheet catalogue provides both optical and near-infrared spectral types when available. For our sample, we adopted optical classifications for L dwarfs and near-infrared types for T and Y dwarfs.} of the selected brown dwarfs. In Figure \ref{jhk-por-te}, we present colour-colour and colour-magnitude diagrams constructed exclusively with magnitudes from the 2MASS catalogue. 
These diagrams reveal a transitional pattern: from spectral types M to approximately T0, the colours become progressively redder (panels (a) to (d)). However, between T0 and T4, the trend reverses towards bluer colours (panel (e)), a behaviour that remains evident in the T5-Y spectral range (panel (f)). This reversal in the colour trend is the phenomenon known as the bluing effect, as mentioned previously. It is important to note that while Figure 2 includes spectral types M0–M5 to provide context and show the continuity of photometric trends, the classification analysis presented in the following sections focuses strictly on brown dwarfs (M6 and later).

This effect is also clearly visible in the colour-magnitude diagrams shown in Figure \ref{jhk-por-te} (panels (g), (h), and (i)). In these panels, it can be observed that starting from the T0-T5 spectral class, all three 2MASS colour indices (J$-$H, H$-$K$_s$, and J$-$K$_s$) become significantly "bluer" (shifting towards lower values). Interestingly, while the colours show this marked transition, the absolute magnitude $M_J$ remains almost constant across these spectral types, creating a near-horizontal distribution in the diagrams for the coldest objects in the sample.

\begin{figure}[!ht]
\centering
\includegraphics[width=\columnwidth]{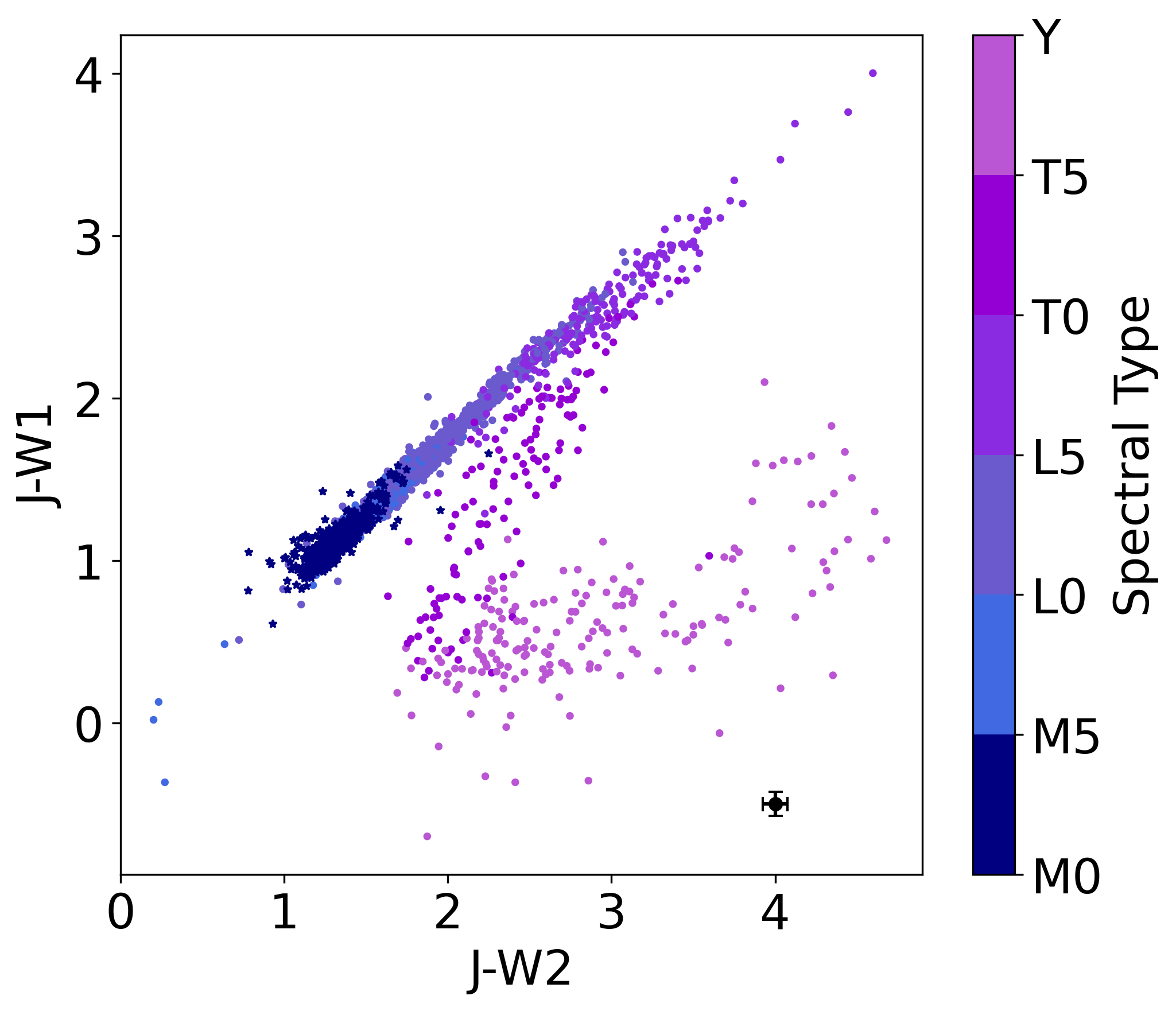}
\caption{Colour-colour diagram (J$-$W1 vs. J$-$W2) for the brown dwarf sample, where 
points are colour-coded by spectral type. The black symbol with error bars in the lower-right corner represents the median uncertainty for all objects in the sample. Note: As indicated in Figure 2, types M0–M5 are shown for contextual continuity but are not part of the spectral classes defined in this work. The T5–Y group is further subdivided into Class 4 (T5–T9) and Class 5 (Y) in our classification scheme (see Table 1).
}
\label{color-color1}
\end{figure}

In Figure \ref{color-color1} we present a colour-colour diagram that includes the magnitudes W1 and W2 of the WISE survey. Incorporating W1 and W2 instead of H and K$_s$ prevents spectral types from entirely overlapping, allowing certain patterns to emerge more clearly. Specifically, for objects with spectral types ranging from M to T0, it is observed that as the spectral type becomes later, both colour indices increase in similar proportions. 
Between spectral types T0 and T5, J$-$W2 remains relatively constant, whereas J$-$W1 decreases significantly, further confirming that these brown dwarfs become bluer.

Lastly, within the T5 to Y range, the trend once again reverses, where later spectral types correlate with increasing colour indices. This ultimate change is anticipated, as at these extremely low temperatures, the methane absorption bands are no longer as prominent as those observed in early to intermediate T-type brown dwarfs \citep{2014PASA...31...43B}. This outcome aligns with theoretical predictions; however, the population of known Y-type brown dwarfs remains quite limited. When \citet{2017ApJ...842..118L} conducted their study, only 24 of these objects were known.  Presently, the UltracoolSheet database lists 49 Y-type brown dwarfs, reflecting a degree of advancement in their identification, yet the sample size is still too restricted to empirically validate this relationship.

\subsection{Multidimensional photometric analysis}

As discussed in the previous sections, the most relevant result so far has been the absence of a clear and well-defined relationship between colour indices and spectral type. This departure from typical trends observed in main-sequence stars significantly complicates the estimation of spectral types using photometric data alone, especially when the analysis is limited to two dimensions (two colour indices). As mentioned in the Introduction, this work leverages modern machine learning algorithms to analyse multidimensional spaces, thereby enabling the simultaneous inclusion of multiple colour indices.

We focus specifically on supervised learning techniques, using magnitudes and colour indices as input features and spectral type as the target variable. To simplify the classification, we grouped the spectral types into six broader classes: Class 0 (M6–M9), Class 1 (L0–L4), Class 2 (L5–L9), Class 3 (T0–T4), Class 4 (T5–T9), and Class 5 (Y). This grouping reduces classification noise between adjacent subtypes and allows the model to capture broader structural patterns in the data. To restrict the classification to the brown dwarf regime, Class 0 was defined to start at spectral type M6, consistent with the inclusion criteria for ultracool dwarfs.

\begin{table}[!ht]
\caption{Number of objects in training and test sets.}
\label{tab_training_test}
\centering                   
\begin{tabular}{c c c}      
\hline\hline               
\text{Spectral class} & \text{Training} & \text{Test} \\
\hline                      
   Class 0 (M6-M9) & 411  & 176 \\    
   Class 1 (L0-L4) & 499  & 214 \\
   Class 2 (L5-L9) &  98  &  42 \\
   Class 3 (T0-T4) &  72  &  31 \\
   Class 4 (T5-T9) & 107  &  42 \\
   Class 5 (Y)    &  19  &   8 \\
\hline                      
\end{tabular}
\end{table}

The dataset was split using a stratified random sampling into a training set (70\%) and a test set (30\%). The training set contains a total of 1206 objects, distributed in spectral classes as shown in Table \ref{tab_training_test}. This distribution ensures that each class is adequately represented for effective model training. All these objects have measured magnitudes in the J, H, and Ks bands from 2MASS and in the W1 and W2 bands from WISE. The W3 and W4 magnitudes were discarded due to their low precision.

In this work, we employ two classification algorithms: Random Forest (RF) and Gaussian Processes (GP). The RF classifier \citep{2001MachL..45....5B} is a widely used supervised learning technique. This algorithm relies on an ensemble of decision trees, where the final prediction is derived from a majority vote among the individual trees. This aggregation strategy enhances accuracy and mitigates overfitting when compared to single-tree models. Once trained on a set of labelled data, the algorithm can then classify new, unseen data into appropriate categories. The hyperparameters of the RF algorithm are: $max\_depth$ (maximum depth of each tree), $min\_samples\_leaf$ (minimum number of samples required at a leaf node), $min\_samples\_split$ (minimum number of samples required to split an internal node), and $n\_estimators$ (number of trees in the forest). The method for the determination of their optimal values is described in Section \ref{feat_eng}.

GP are a non-parametric supervised learning method used for both regression and classification problems. Their primary advantages are the ability to make probabilistic predictions and great versatility, owing to their support for various kernel types. The kernel used in this work was the 'Radial Basis Function (RBF)', parameterized by a length-scale that can be either a scalar or a vector matching the input dimensionality. In this case, it was a vector with a dimension equal to the number of classes, whose values were determined and optimized by the algorithm during the training process. Since the RBF kernel depends on the distance between points, proper data normalization is crucial to prevent one variable from dominating over the others due to differences in their magnitudes. For this reason, the data was preprocessed using the \textit{StandardScaler} method from the \textit{sklearn.preprocessing} library\footnote{https://scikit-learn.org/stable/modules/preprocessing.html}. This method performs standard normalization according to the relation $X_{norm} = \frac{X - \mu}{\sigma}$, transforming each feature to have a mean of 0 and a standard deviation of 1. This ensures that all variables contribute equally to the model. We employ the RF and GP algorithms from scikit-learn \citep{2011JMLR...12.2825P}. 

\subsection{Feature engineering and model evaluation} \label{feat_eng}

\begin{figure}[!ht]
    \centering
    \subfloat[]{%
    \includegraphics[width=0.65\linewidth]{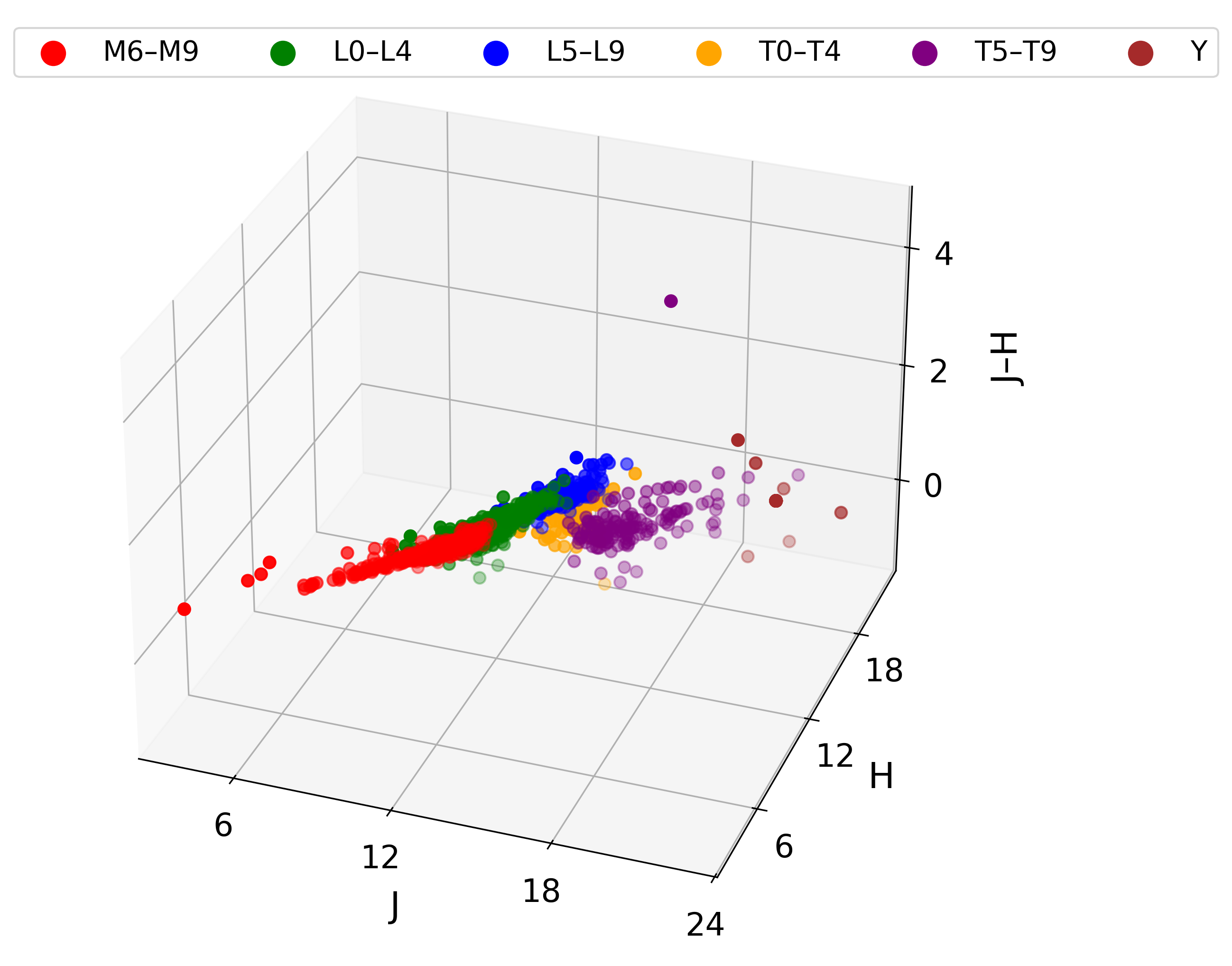}
    }
    \par\vspace{0.2cm}
    \subfloat[]{%
   \includegraphics[width=0.65\linewidth]{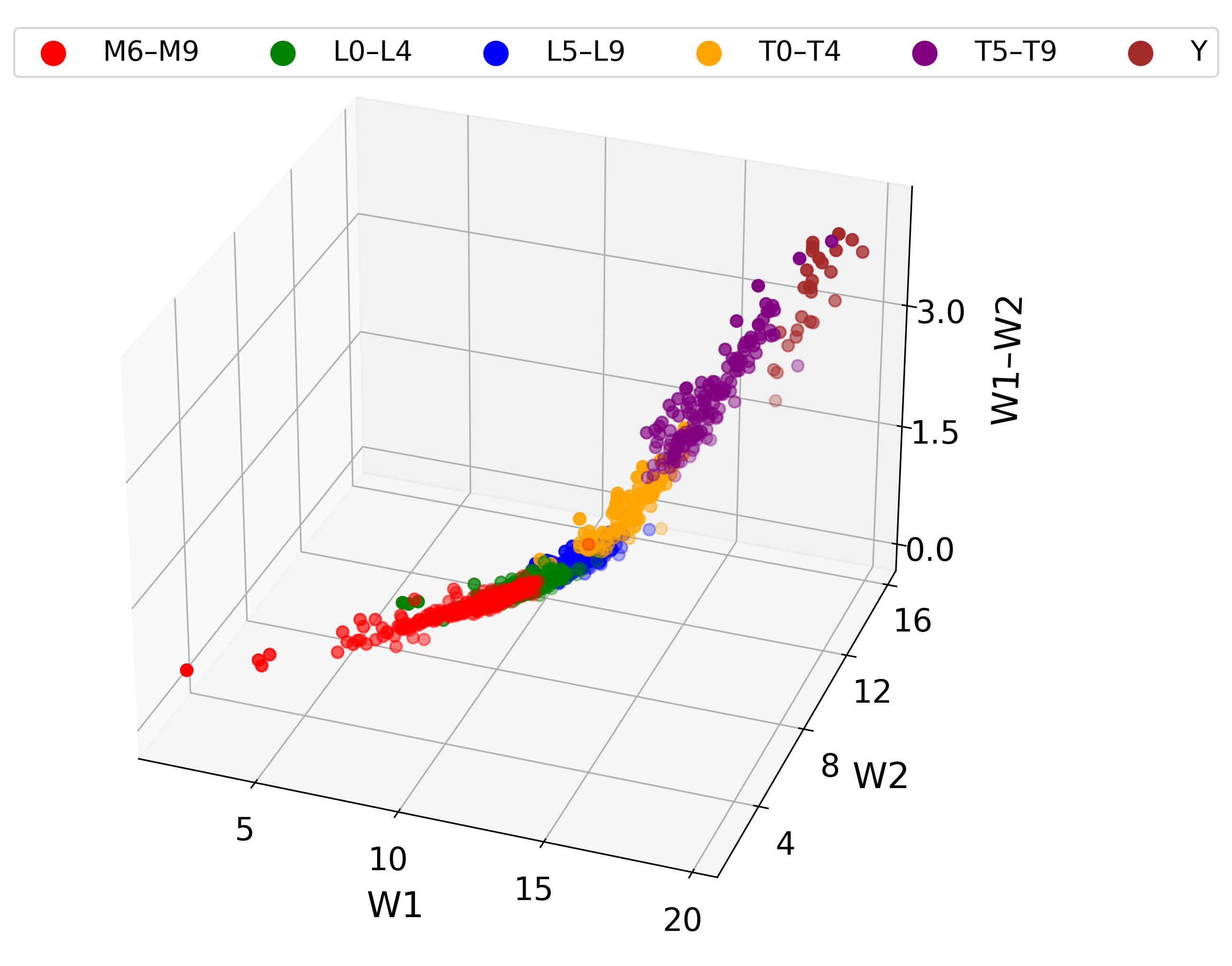}
    }
    \caption{Panel (a): J vs H vs J$-$H, panel (b): W1 vs W2 vs W1$-$W2, where J, H, W1 and W2 are absolute magnitudes. In both plots, the different colours represent the classes associated with spectral types as determined in the literature.}
    \label{3d_ultracool}
\end{figure}

The feature selection process was performed for the RF classifier. Once the optimal combination was identified, the same set of features was used for the GP model to ensure a consistent comparison. We began examining two three-dimensional spaces, as shown in Figure \ref{3d_ultracool}. The top plot displays absolute magnitudes and colours from the 2MASS catalogue, while the bottom plot uses photometric data from the WISE catalogue. In the
case of the 2MASS data, the J$-$H colour separated spectral classes
better than the J$-$K colour, exhibiting, as well, less dispersion
(see Figure \ref{color_te(2)}).

In these diagrams, we observe that incorporating an additional dimension leads to greater separation between spectral types compared to two-dimensional diagrams. This separation is particularly noticeable in the WISE colour indices, whereas in the 2MASS data, some overlap between
spectral types still exists.

\begin{figure}[!ht]
\centering
\includegraphics[width=\columnwidth]{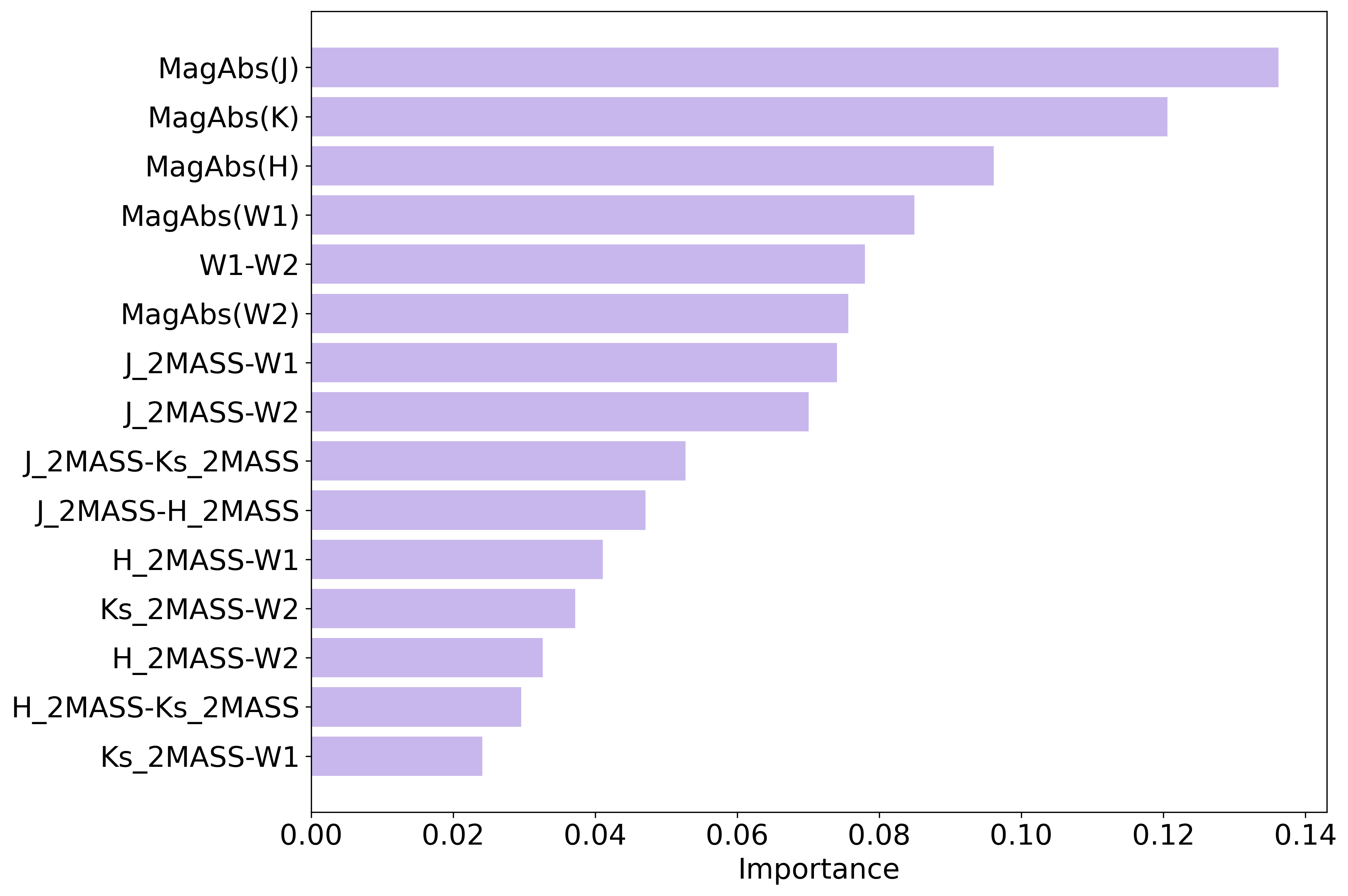}
\caption{The importance assigned by the Random Forest algorithm to each of the available features in the sample.}
\label{imp_rf}
\end{figure}

\begin{figure}[!ht]
\centering
\includegraphics[width=\columnwidth]{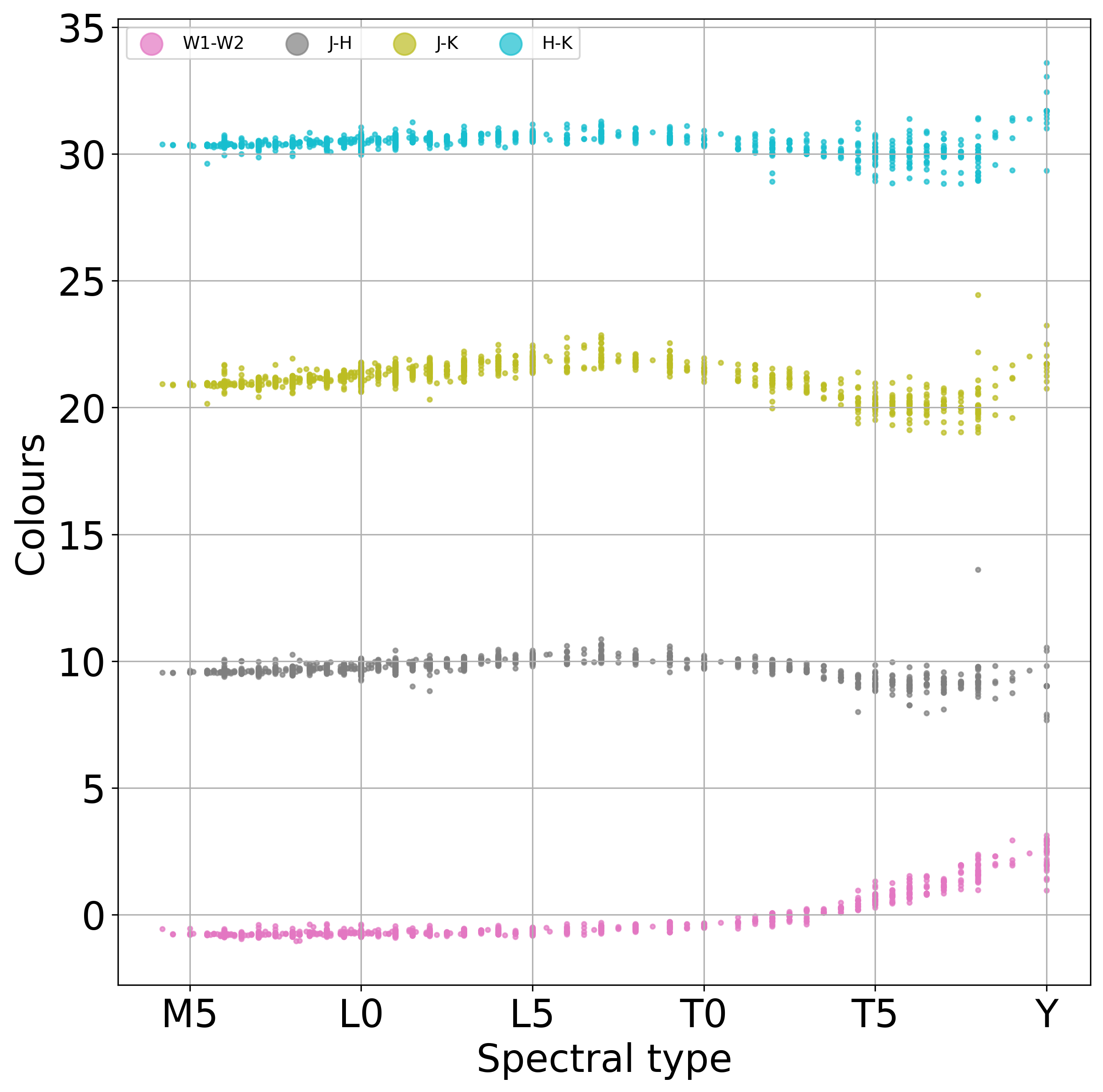}
\caption{2MASS and WISE colour indices shown separately as a function of spectral type. Each of the indices has been vertically shifted for better visualization. Typical photometric uncertainties are on the order of $\sim$ 0.1 mag for the 2MASS colours and $\sim$ 0.02 mag for the WISE colour; these are not shown in the figure for clarity.} 
\label{color_te(2)}
\end{figure}

\begin{figure}[!ht]
\centering
\includegraphics[width=\columnwidth]{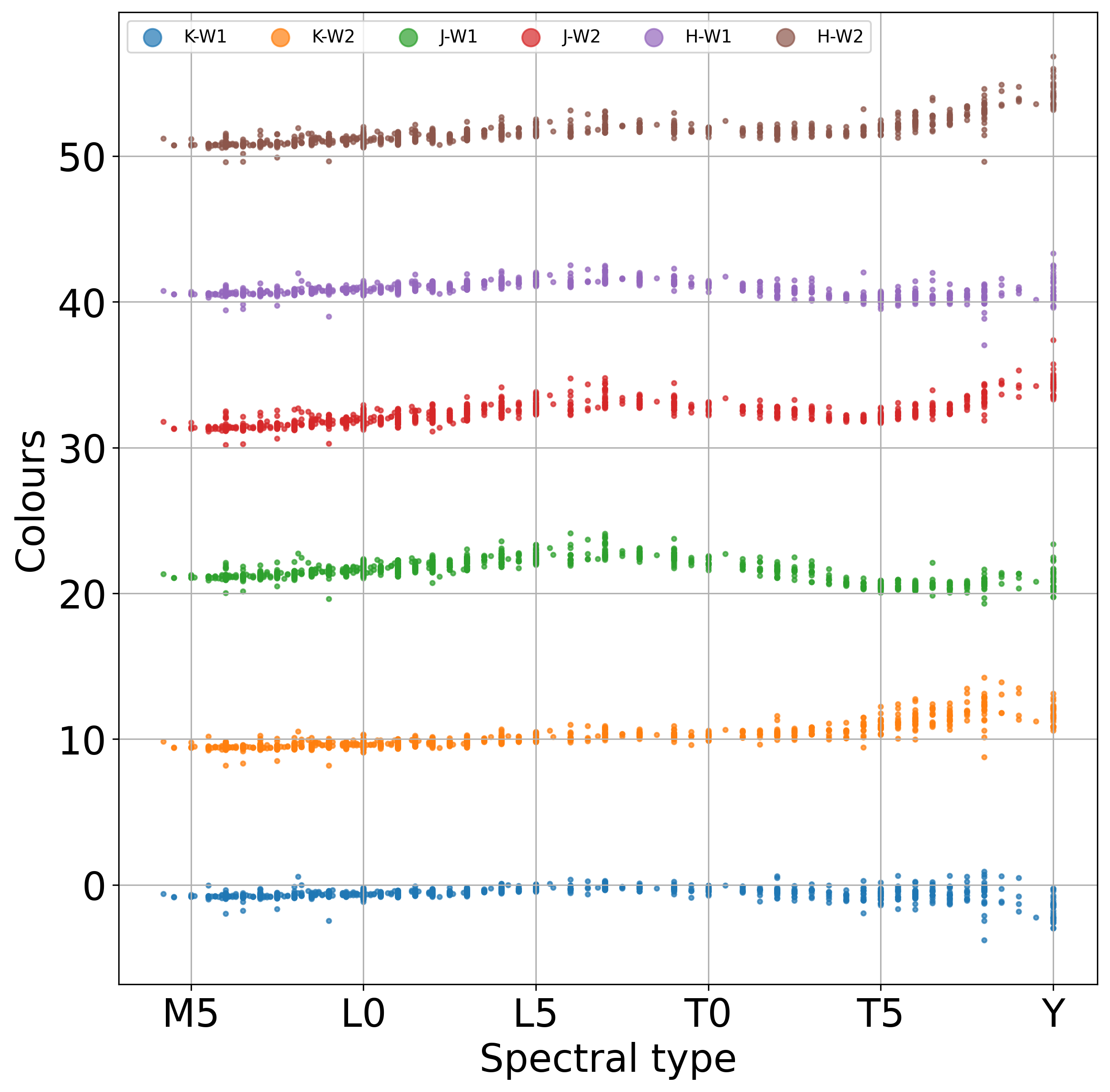}
\caption{Combined 2MASS + WISE colour indices as a function of spectral type. Each of the indices has been vertically shifted for better visualization. Typical photometric uncertainties for all colours are on the order of $\sim$ 0.07 mag and are not shown in the figure for clarity.}  
\label{color_te(1)}
\end{figure}

In addition, we analysed, on the one hand, the importance of each feature for the algorithm (Figure \ref{imp_rf}) and, on the other hand, the colour trends as a function of the spectral type (Figures \ref{color_te(2)} and \ref{color_te(1)}). From Figure \ref{imp_rf}, we can see that the absolute magnitudes J, H, K, and W1 are the most relevant features for the classifier's performance.

Furthermore, an examination of the trend of various colour indices as a function of spectral type in Figure \ref{color_te(2)} shows that W1$-$W2 exhibits a distinct increasing trend for spectral types later than T0. Conversely, the J$-$H and J$-$K colours exhibit an initial increasing trend, peaking around spectral types L5-T0, followed by a decrease toward later spectral types. In contrast, the colour H$-$K displays no variation across spectral types, thus suggesting W1$-$W2, J$-$H, and J$-$K as the most promising candidates. Subsequent classifier performance tests, both with and without the J$-$K colour, demonstrated a slight improvement upon its exclusion\footnote{Tests performed excluding the J-K colour marginally improved the F1-Score from 0.84 to 0.85.  Likewise, omitting the K-band magnitude resulted in an increase to 0.86.}. Consequently, J$-$K was discarded, with only the W1$-$W2 and J$-$H colours retained.

We note that, in particular, near-infrared colours (J$-$H), (J$-$K) and (H$-$K) show a significant scatter of about 1 magnitude. \citet{2023ApJ...954L...6S} provide direct observational evidence that the atmospheres of mid-L dwarfs exhibit a latitudinal dependence of dust cloud opacity, where equatorial regions are significantly cloudier and more opaque than the more transparent polar regions \citep{2017ApJ...842...78V}. This distribution implies that an object's appearance is heavily influenced by its viewing geometry; specifically, L dwarfs viewed equator-on appear much redder than those viewed at higher latitudes, with colour variations exceeding 1 magnitude in the near-infrared (J$-$K)  and reaching approximately 0.2 magnitudes in the mid-infrared (W1$-$W2).

In Figure \ref{color_te(1)}, we show colour indices formed by a combination of magnitudes from 2MASS and WISE. From these colours, we see that K$-$W1 and K$-$W2 show very little variation across all spectral types, so these were the first colours to be discarded. Among the remaining four colours (J$-$W1, J$-$W2, H$-$W1, and H$-$W2), we observe a trend similar to that seen in J$-$H and J$-$K, with J$-$W1 showing the most noticeable changes across spectral types, followed by J$-$W2.

Using the training set, we performed a 5-fold cross-validation to determine the optimal hyperparameters. The data set is divided into five folds; in each iteration, four folds are used for training and one for validation. This is repeated five times, ensuring that each fold serves as the validation set once. The performance of the model is averaged across all iterations to select the best set of hyperparameters.

In summary, the selected features used for both the GP and RF algorithms were the absolute magnitudes J, H, and W1, and the colours W1$-$W2, J$-$W1, J$-$W2, and J$-$H. With this feature set, the RF model was configured with the following hyperparameters: $max\_depth: 8$, $min\_samples\_leaf: 2$, $min\_samples\_split: 2$, and $n\_estimators: 100$. The GP model, on the other hand, utilized a Radial Basis Function (RBF) kernel with a trainable length scale vector, the final values being (7.02, 2.43, 5.9, 3.88, 3.49, 7.11).

Finally, evaluation metrics give a measure of how well a model performs. One of the evaluation metrics used was the F1-Score\footnote{https://scikit-learn.org/stable/modules/generated/sklearn.metrics.f1\_score.html}. The F1-Score is the harmonic mean of precision and recall, where an F1 score reaches its best value at 1 and its worst score at 0. Precision measures the proportion of correctly predicted positive instances among all predicted positives, and recall measures the proportion of correctly predicted positives among all actual positive instances. This metric is particularly useful when dealing with imbalanced datasets, given its ability to balance the trade-off between precision and recall. Furthermore, in addition to the F1-Score, we utilized confusion matrices to compare our classification results with those reported in the literature.
\section{Results}

\subsection{Performance of the classifiers}

Table \ref{tab_evaluation_metrics} presents the performance metrics for each classifier. Both the RF and the GP classifiers achieved high scores in all three metrics. Although the GP classifier exhibited marginally higher values, both models demonstrated comparable and consistently robust performance. This indicates that  either can be confidently employed for this classification task.

\begin{table}[!ht]
\caption{Evaluation metrics for the RF and GP classifiers.}  
\centering                        
\begin{tabular}{c c c c}      
\hline\hline               
\text{Algorithm} & \text{Precision} & \text{Recall} & \text{F1-Score} \\         
\hline                      
   Random Forest (RF) & 0.88  & 0.84  & 0.86 \\    
   Gaussian Processes (GP) & 0.89 & 0.86 & 0.87  \\
\hline                                  
\end{tabular}
\label{tab_evaluation_metrics}    
\end{table}

\begin{figure*}[!ht]
\centering
\includegraphics[width=0.48\textwidth]{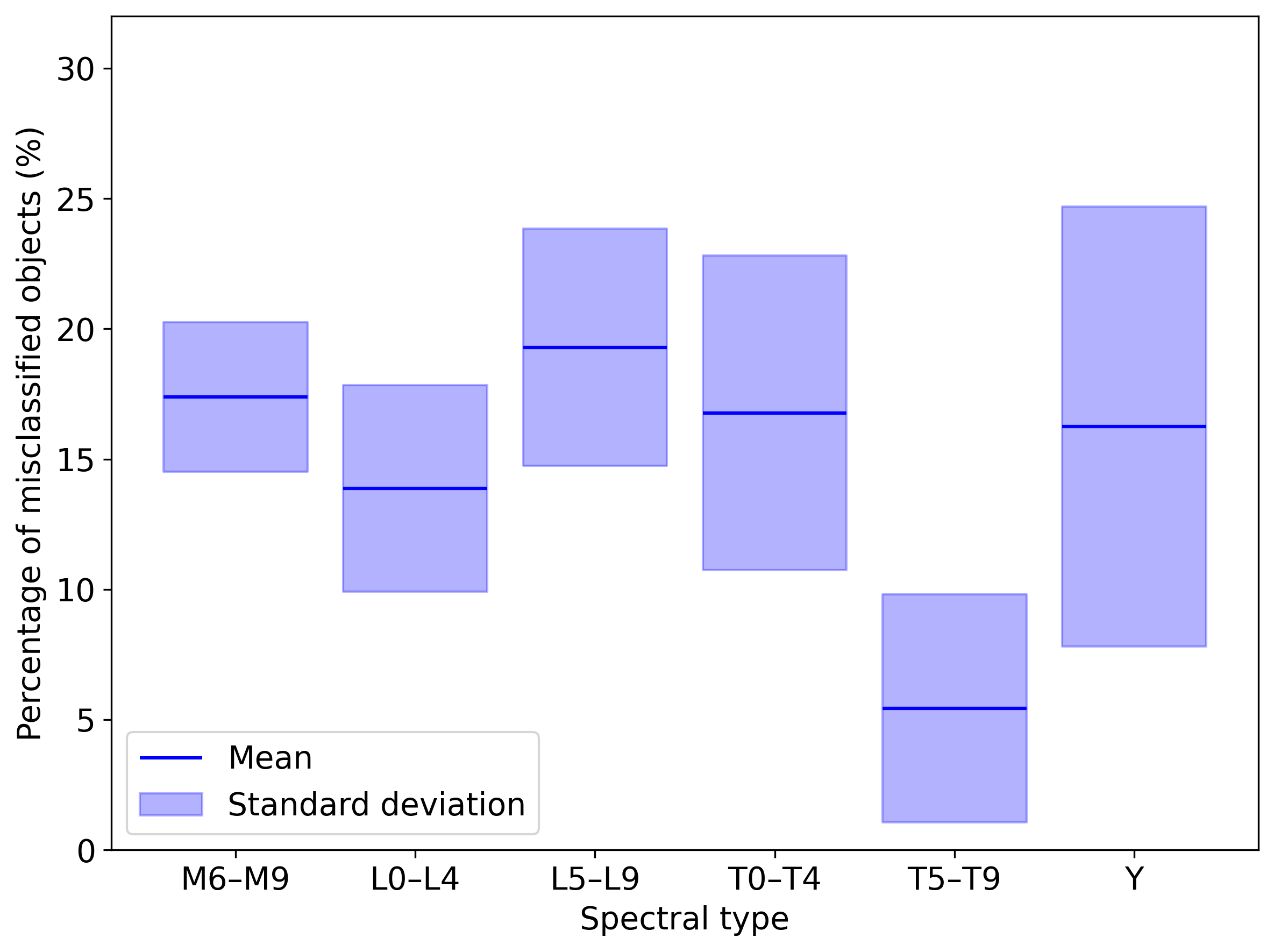}
\includegraphics[width=0.48\textwidth]{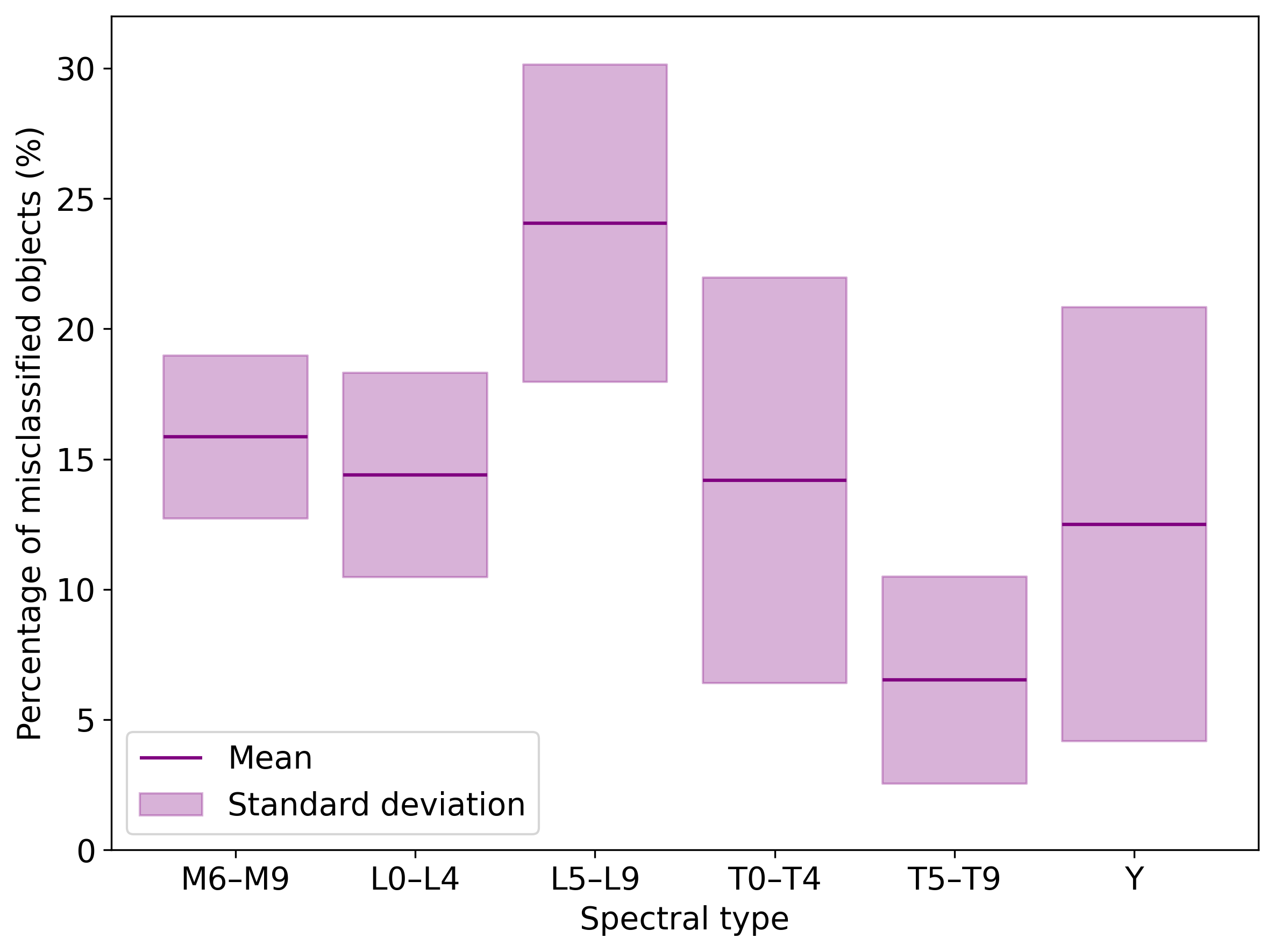}
\caption{Percentage of misclassified objects per spectral type range using the Random Forest (left) and Gaussian Processes (right) classifiers. In each panel, the solid line represents the average from 10 runs, while the shaded area indicates the standard deviation (blue for RF, pink for GP).} 
\label{fig:promedios_rf_gp}
\end{figure*}

To obtain a more accurate estimate of each classifier's per-class performance, we executed 10 iterations of each algorithm, employing distinct random states for the split of the sample into training/testing sets. This approach serves to eliminate any bias that might be associated with the selected dataset. In each run, we calculated the percentage of misclassified objects per class and then derived an average percentage and its standard deviation. Figure \ref{fig:promedios_rf_gp} shows these results for the RF and GP classifier. We highlight that the limited number of Y-type brown dwarfs (27 objects in our sample) leads to a high classification error for this class. Therefore, the results should be considered as a first approximation. We note, however, that the uncertainty for the Y dwarfs is similar to those of early-T types, which are more numerous. This may be related to the onset of the reversal in the colour trend within the T0–T4 class (see Figure \ref{jhk-por-te}, panel (e)).

Classes 1 (M6–M9) and 2 (L0–L4) show similar uncertainties, which are relatively small compared to classes 3 (L5–L9) and 4 (T0–T4). This is likely due to the positive correlation between colour and spectral type in these classes. Furthermore, we observe that both algorithms show a marked decrease in misclassified objects for class T5–T9. This improvement can be attributed to the fact that, from spectral type T5 onwards, 2MASS magnitudes are no longer affected by the bluing effect (Figure \ref{color-color1}). In addition, L and early-T type dwarfs show larger colour scatter, as their atmospheres are dominated by thick silicate clouds. In contrast, from T5 onwards, brown dwarfs exhibit less colour scatter because these silicate clouds begin to sediment out of the visible photosphere as the objects cool. This process results in atmospheres that are less affected by the viewing-angle-dependent opacity seen in cloud-laden L and early-T type dwarfs \citep{2017ApJ...842...78V, 2023ApJ...954L...6S}. In other words, the viewing-angle effect is mitigated as silicate clouds sediment out of the photosphere. Consequently, near-infrared colours follow a monotonic trend where redder colours consistently correspond to later spectral types. This well-defined colour–spectral type relationship facilitates the accurate classification of these objects.

\begin{figure}[!ht]
    \centering
    \includegraphics[width=\columnwidth]{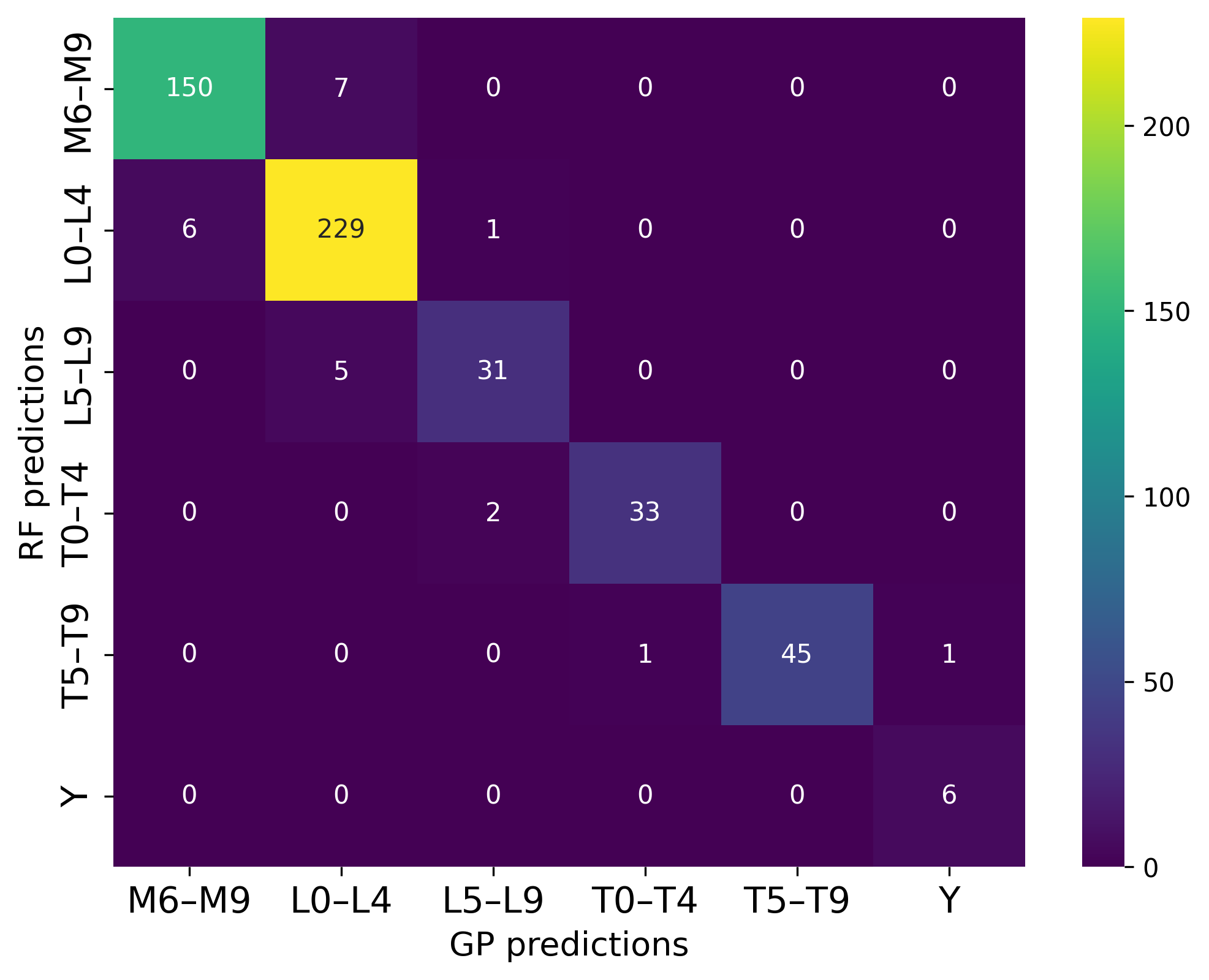}
    \caption{Confusion matrix between the spectral types predicted by RF and GP. The colour scale and the number in each matrix element represent the number of objects.}
    \label{rf_vs_gp}
\end{figure}

As a second step in gauging the algorithms, Figure \ref{rf_vs_gp} presents a confusion matrix that compares the spectral types predicted by both algorithms. 
This shows a strong agreement with their predictions. However, there is some confusion between classes 0 and 1, which correspond to spectral types M6–M9 and L0–L4 respectively, that affects 3.4\% of those objects. Another notable trend is the decreasing number of objects as we move toward later spectral types. This is because such objects are very faint, making them difficult to detect and classify. Specifically, we see that there are only 7 Y spectral-type objects in the test set, which aligns with the large dispersion in the misclassification rate for this class.

\subsection{Comparison with previous studies}

In this section, we compare our spectral type determinations with results from three significant studies in the field to validate the robustness of our model. These studies utilized different methodologies for identifying and classifying ultracool dwarfs:

\citet{2015A&A...574A..78S}: This study employed a photo-type method, a specialized photometric technique using 8-band photometry from SDSS (i, z), UKIDSS (Y, J, H, K), and WISE (W1, W2). Classification was performed by comparing measured magnitudes against spectral templates derived from polynomial fits to the colours of spectroscopically confirmed dwarfs. Using a $\chi^2$ minimization scheme, the authors identified 1,361 brown dwarfs ranging from L0 to T8. For our analysis, we compared our spectral classes with the photometrically derived types for 150 objects from their sample, all of which fall within the L0–L7 spectral range.
The authors report that their method achieves precision within one spectral subtype (RMS error), a result validated via three independent methods: comparison with documented sources, SpeX instrument follow-up of a small sample, and Monte Carlo simulations.

\citet{2024AJ....168..211B}: The authors identified 118 new ultracool dwarf candidates by applying the SMDET convolutional neural network to WISE time-series images \citep{2018AJ....156...69M}. These candidates were vetted by citizen scientists and classified using the photo-type estimation method \citep{2015A&A...574A..78S}, subsequently validated through near-infrared spectroscopy of two objects. Although their study spans the M4–T7 spectral range, the 20 objects shared with our sample are concentrated within the M6–T2 interval. Furthermore, even if photometric estimations are generally accurate within one subtype, the authors note a systematic bias toward earlier classifications for late-type T-dwarfs --objects that lie beyond the scope of our comparative analysis.

\citet{2025A&A...695A.195K}: This work conducted a systematic search for high proper motion objects using the unTimely WISE catalogue \citep{2023AJ....165...36M}. Their pipeline integrated a motion detection algorithm with an iterative Random Forest binary classifier and visual inspection. Spectral types, ranging from M7 to Y0, were estimated using the VOSA tool with BT-Settl model atmospheres \citep{2008A&A...492..277B}.
In this study, the precision of the spectral type determination is variable rather than fixed; the associated error depends on the volume of photometric data available and the specific classification method applied to each object. Consequently, final classifications are often reported as intervals (e.g., T7–T8), reflecting an uncertainty that generally remains within two subtypes. Our study overlaps with this sample specifically in 75 objects within the M5-T6 spectral range.

\begin{figure}[!ht]
    \centering
    \includegraphics[width=\columnwidth]{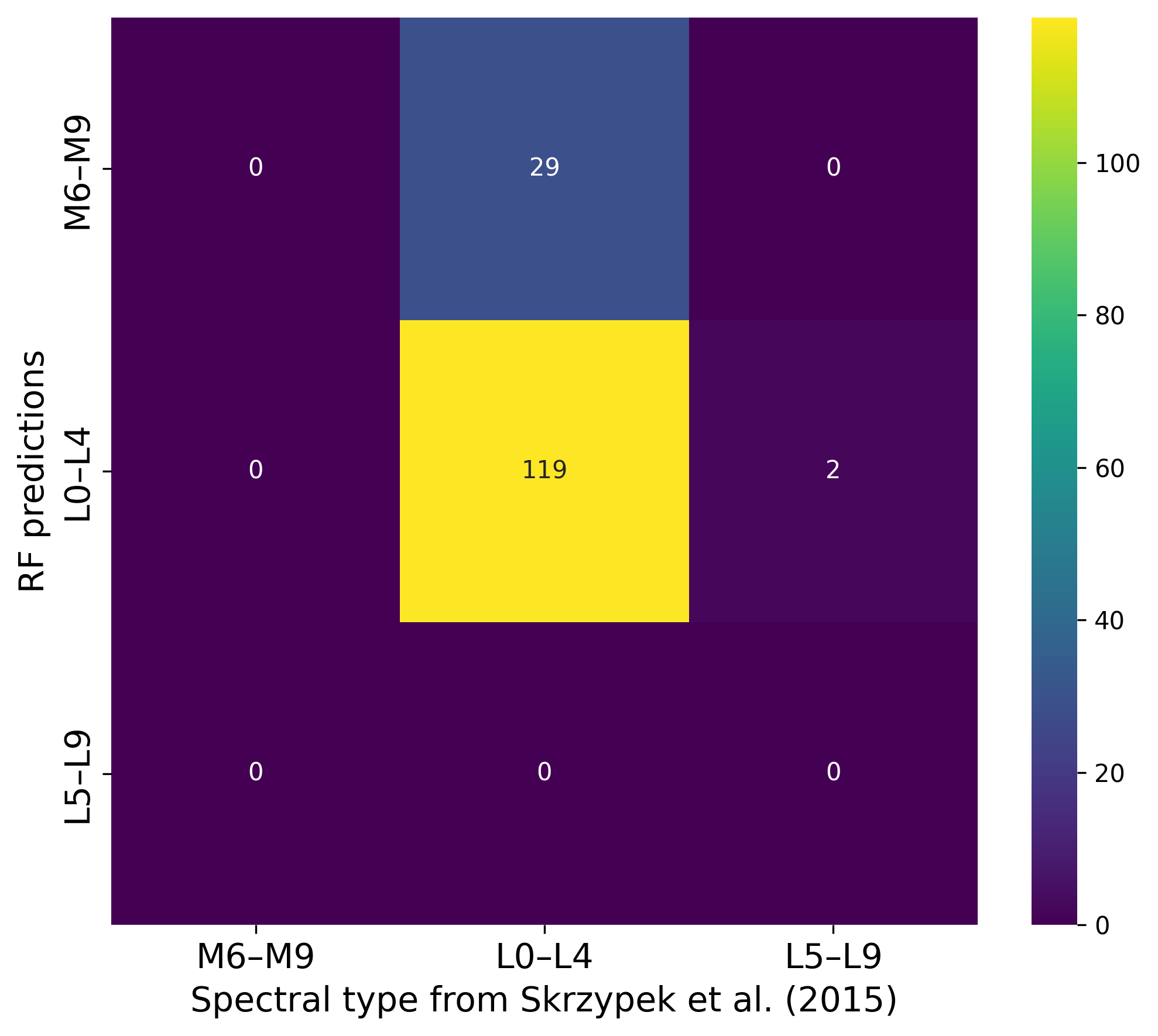}
    \caption{Confusion matrix comparing the spectral types predicted by our Random Forest model with those estimated by \citet{2015A&A...574A..78S} using polynomial fits of colour indices.}
    \label{studies_rf}
\end{figure}

All comparisons are restricted to the subsets of sources from previous studies that met the required photometric features for our model; hence, these comparisons span narrower spectral ranges than those covered in the original studies.

\begin{figure*}
\centering
\includegraphics[width=0.48\linewidth]{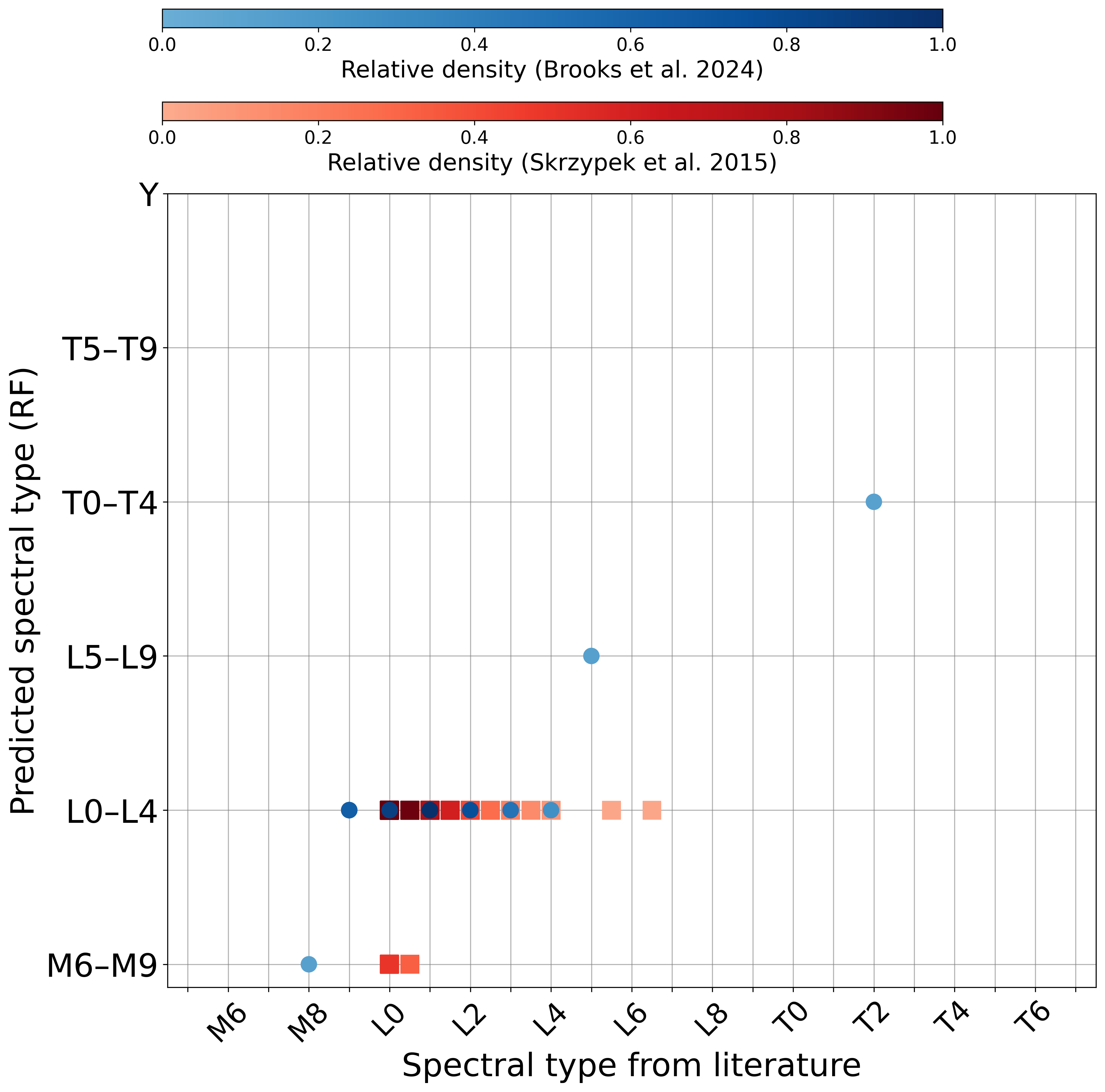}
\includegraphics[width=0.48\linewidth]{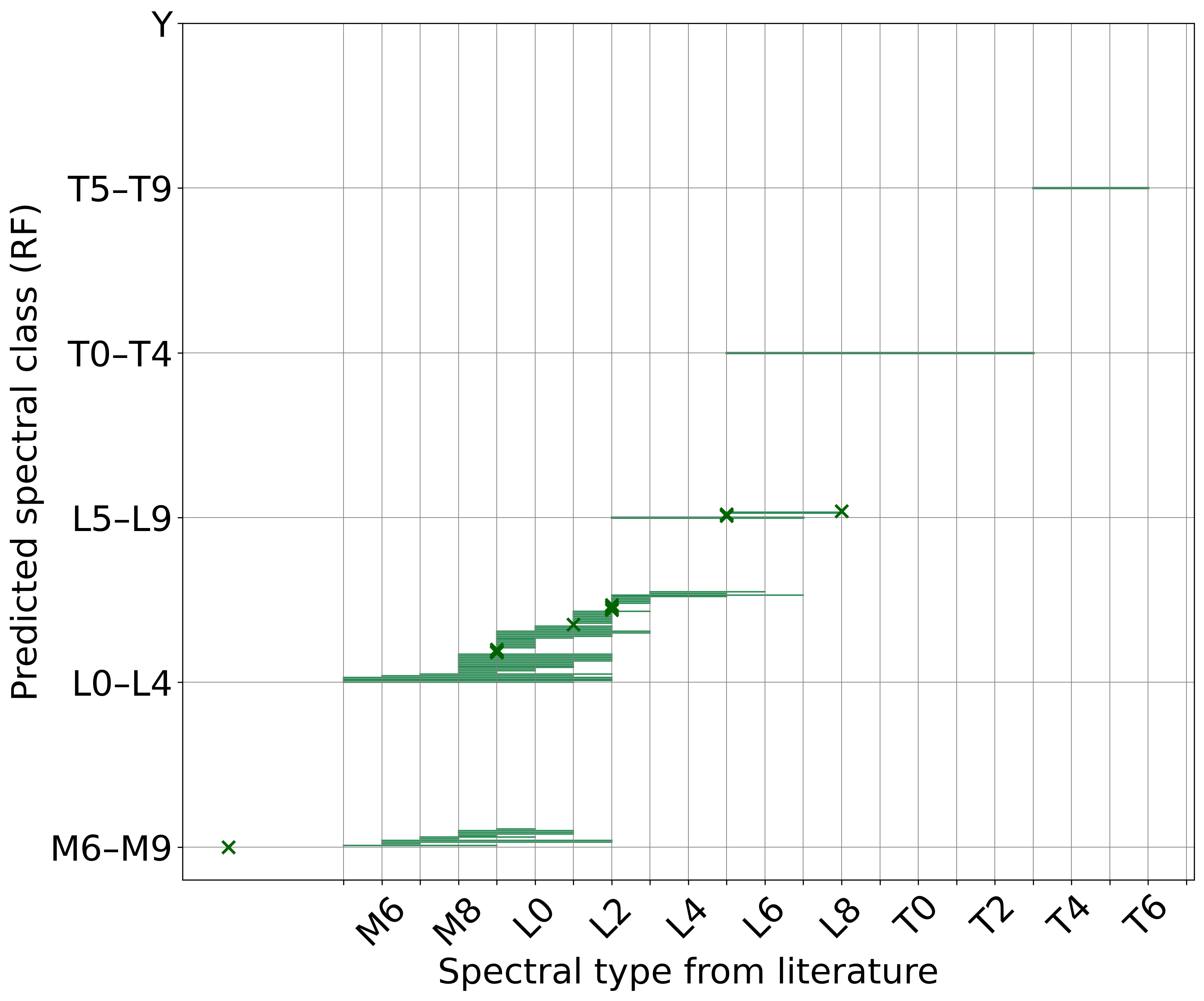}
\caption{Comparison of spectral types predicted by our Random Forest model and those reported in the literature. Left hand panel:  Comparison with spectral types assigned by \cite{2015A&A...574A..78S} (red) and \cite{2024AJ....168..211B} (blue).  Darker colours indicate a higher number of objects per spectral subtype. Right hand panel: Spectral subtype estimates by \cite{2025A&A...695A.195K} are shown as crosses for individual subtypes, while horizontal lines indicate the full spectral type intervals assigned to each object. These lines are vertically offset for clarity and to prevent overlap.}
\label{comparacion}
\end{figure*}

The results, illustrated in the confusion matrix (Figure \ref{studies_rf}) and the comparison plots (Figure \ref{comparacion}), demonstrate strong overall agreement. Regarding the samples from \citet{2015A&A...574A..78S} and \citet{2024AJ....168..211B}, we found an 80\% agreement between their photo-type classifications and our machine learning-based predictions.

Thirty-one objects from a set of 150 have a photo-type classification that falls outside the spectral class assigned by the RF or GP classifiers: 29 are classified as M9 via photo-typing but fall into the L0–L4 class according to the RF model, while the remaining two (L5.5 and L6.5) are also placed in the L0–L4 class. These small differences, affecting only 20\% of the shared objects, indicate high consistency between the datasets. Figure~\ref{studies_rf} presents the confusion matrix for the comparison with \citet{2015A&A...574A..78S}. Given the similar performance of the RF and GP models, only the RF results are shown. Comparison with \citet{2024AJ....168..211B} yielded a nearly identical confusion matrix, further confirming the robustness of our approach.

Comparing our results with those of \citet{2025A&A...695A.195K} presented a specific challenge, as their spectral types are reported as broad ranges (e.g., L2–L7)\footnote{Unlike our spectral classes, which correspond to fixed bins (e.g., M6–M9, L0–L4), \citet{2025A&A...695A.195K} determine spectral types within variable-sized intervals for each object.}. While this prevented a direct comparison via a confusion matrix, visual inspection of Figure \ref{comparacion} confirms that our predicted classes typically overlap with their reported intervals. Among the 75 objects in common with \citet{2025A&A...695A.195K}, 31 have spectral types or intervals entirely contained within our spectral classes, while 39 show overlapping intervals. Four objects with an M9 type or M8–M9 interval were assigned to the L0–L4 class. Only one object shows a significant discrepancy, classified as M2 while being assigned to the M6–M9 class; however, this is expected because spectral types as early as M2 were not included in our training set, leading the algorithm to assign the object to the earliest available spectral class.

In summary, comparisons with \citet{2015A&A...574A..78S}, \citet{2024AJ....168..211B}, and \citet{2025A&A...695A.195K} indicate a good level of consistency despite differences in binning schemes and the diverse methodologies employed—ranging from polynomial fits and spectral templates to convolutional neural networks and VOSA-based atmospheric modeling.

\subsection{Classification of brown dwarfs detected by direct imaging}

Once the algorithms were trained and their effectiveness was confirmed using the test set and by comparing the results with previous studies, we applied them to estimate the spectral types of brown dwarfs detected via direct imaging listed in the Encyclopaedia of Exoplanetary Systems\footnote{https://exoplanet.eu/home/}. This sample includes 570 isolated brown dwarfs, selected by applying a mass cut between 13 and 80 $M_{\rm Jup}$.

To obtain WISE and 2MASS bands magnitudes, we performed a cross-match with the full WISE catalogue using the right ascension ($\alpha$) and declination ($\delta$) coordinates provided by the Encyclopaedia of Exoplanetary Systems. We searched within a radius of 5 arcseconds around each coordinate to identify the corresponding brown dwarf. We found that out of 570 brown dwarfs, 427 were successfully cross-matched with the WISE and 2MASS catalogues. However, not all of these objects had measurements for every required magnitude, which prevented the use of all classifier features for every object. Consequently, we applied a second filter, selecting only those brown dwarfs with available data for all necessary features. This resulted in a final sample of 217 brown dwarfs. None of the objects in this final sample is present in the training set, ensuring an independent evaluation of the classifier's performance.

A more detailed search revealed that out of the 217 brown dwarfs, 196 had a spectral type previously determined in the literature, while 21 did not. Therefore, we used the 196 brown dwarfs with known spectral types to provide further validation of the performance of our classifier.

\begin{figure}[!ht]
\centering
\includegraphics[width=\columnwidth]{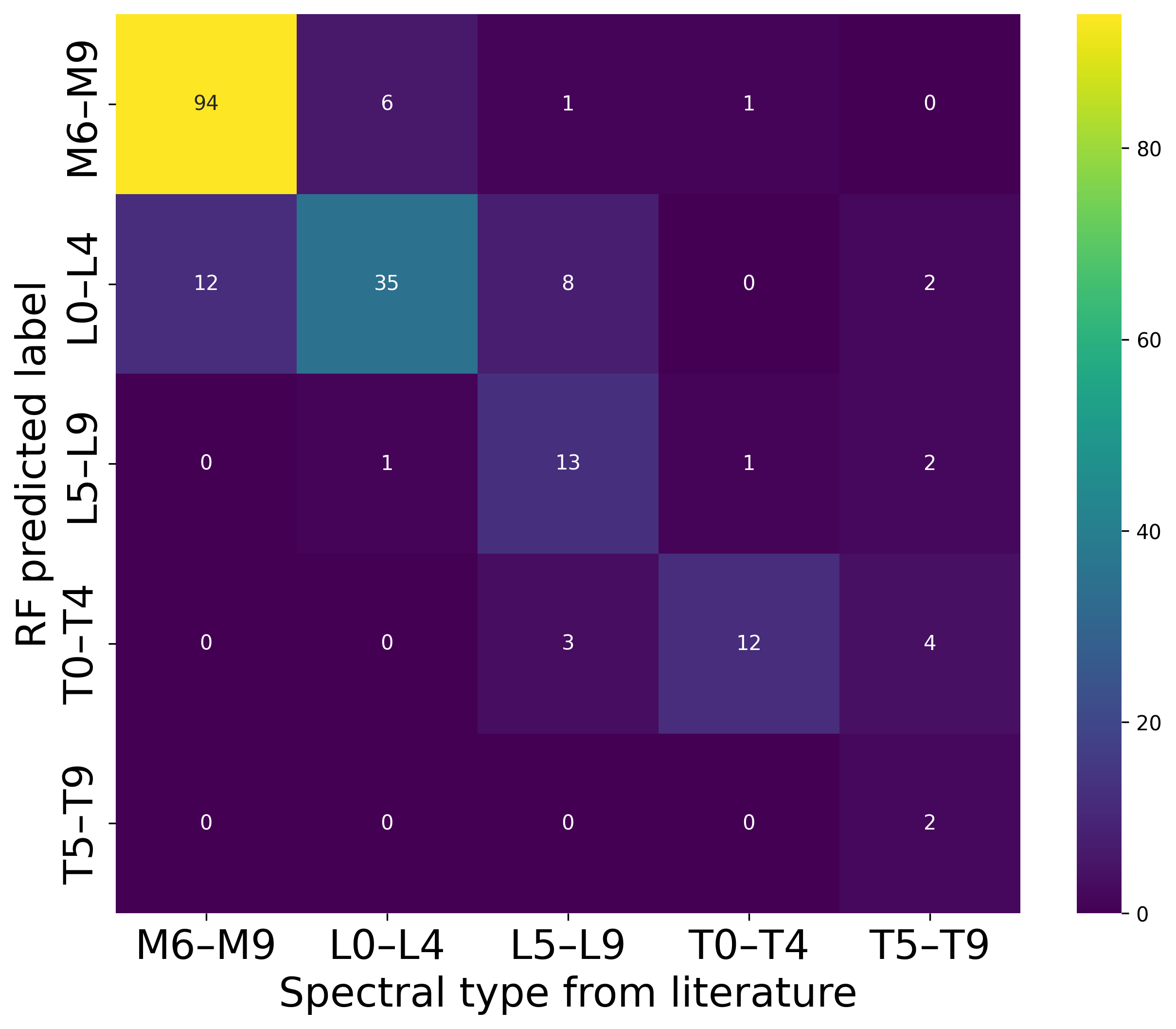}
\caption{Confusion matrix comparing the literature spectral types and those predicted by the RF classifier for isolated brown dwarfs. The colour scale and the number in each matrix element represent the number of objects.}
\label{matriz_conf_rf_nan}
\end{figure}

\begin{figure}[!ht]
\centering
\includegraphics[width=\columnwidth]{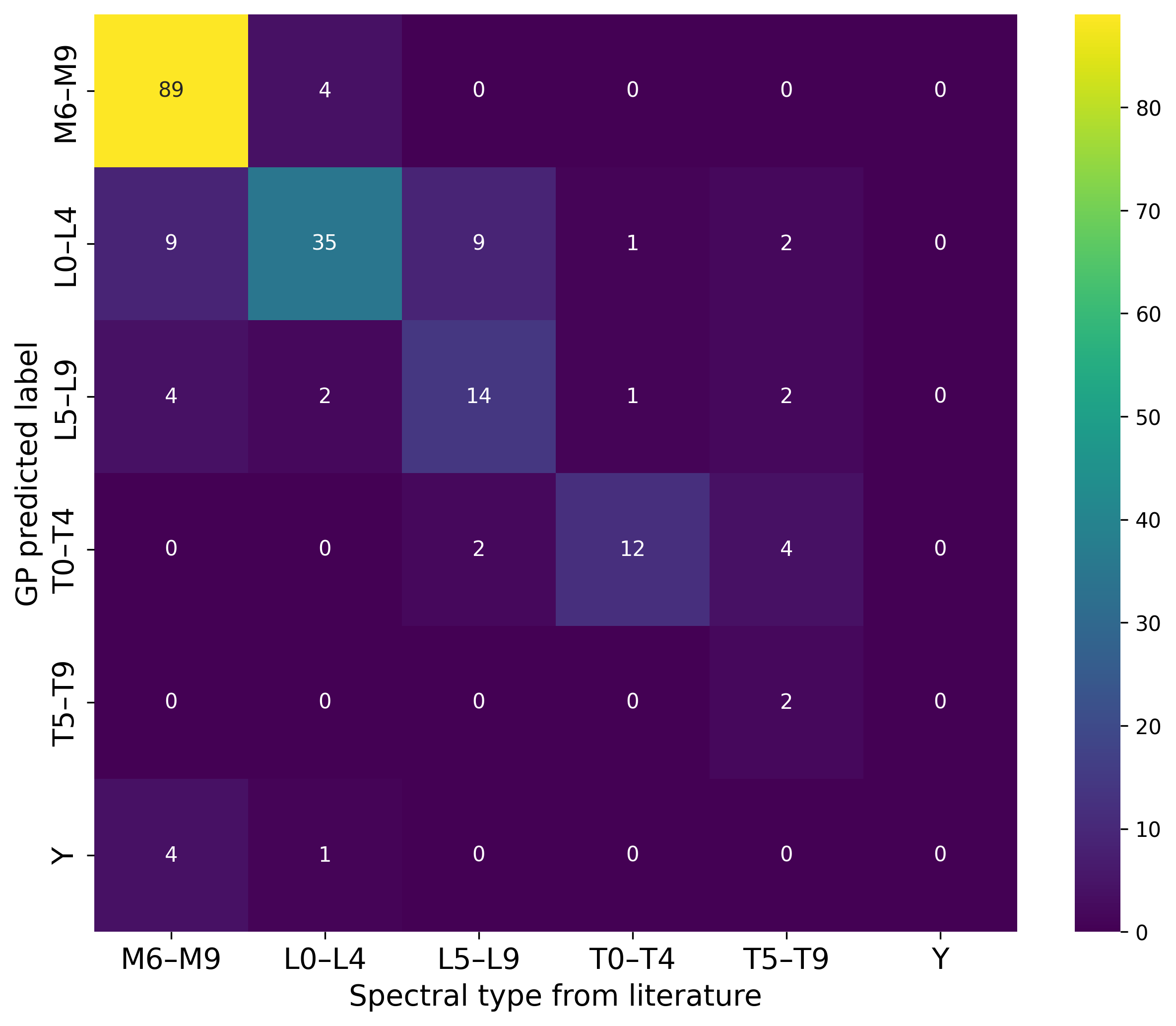}
\caption{Confusion matrix comparing the literature spectral types and those predicted by the GP classifier for isolated brown dwarfs. The colour scale and the number in each matrix element represent the number of objects. }
\label{matriz_conf_gp_nan}
\end{figure}

The application of the RF and GP classifiers resulted in approximately 80$\%$ of the sample correctly classified, as illustrated in Figure \ref{matriz_conf_rf_nan} and Figure \ref{matriz_conf_gp_nan}. Importantly, only 3$\%$ (RF) and 7$\%$ (GP) of the sample were assigned to a spectral class that is not adjacent to the corresponding class according to the literature, suggesting that the classifier successfully captured the relevant features required for a reasonable prediction of the spectral class.

\begin{figure}[!ht]
\centering
\includegraphics[width=\columnwidth]{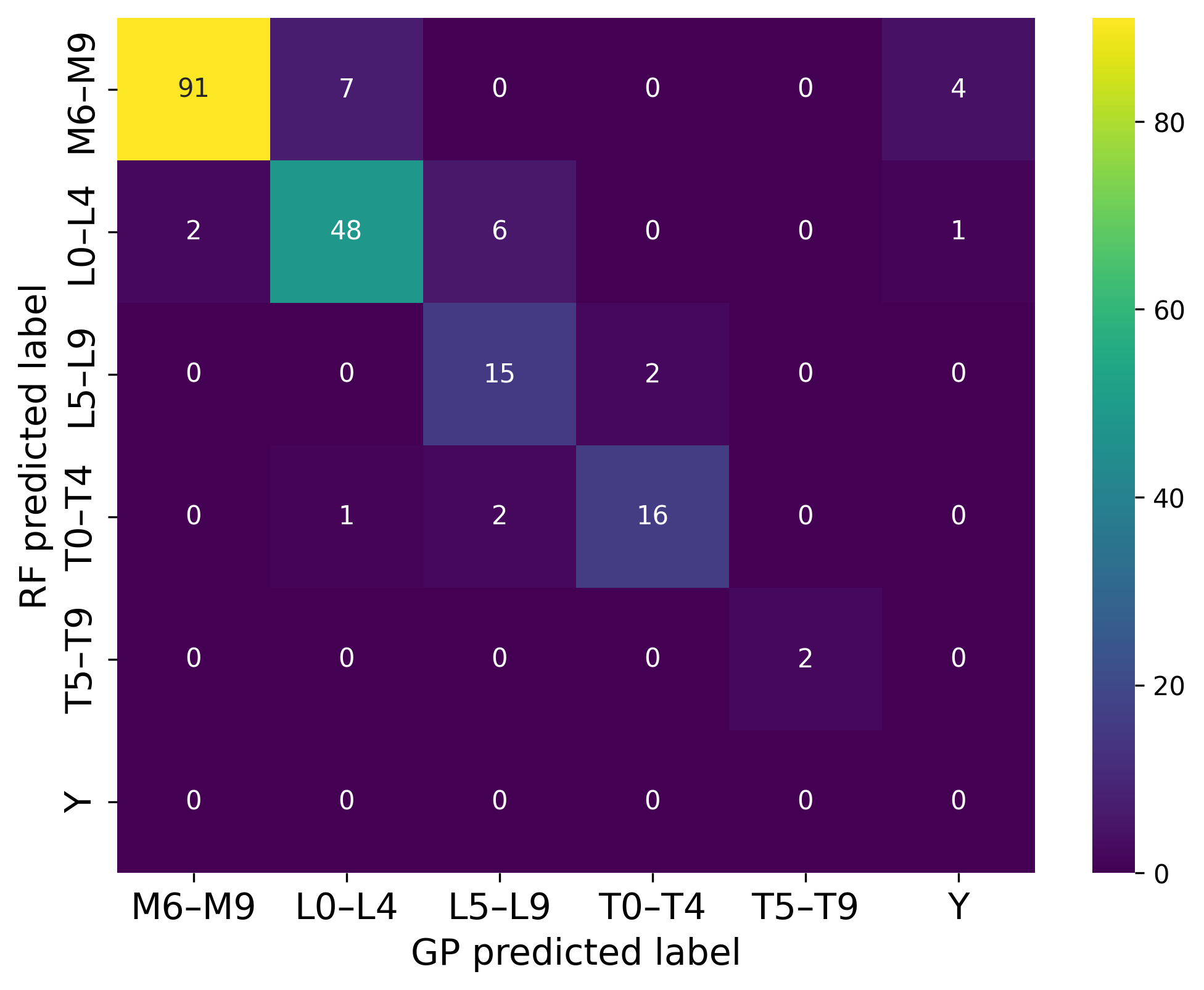}
\caption{Confusion matrix comparing the spectral types predicted by RF and those predicted by the GP classifier for isolated brown dwarfs detected via direct imaging. The colour scale and the number in each matrix element represent the number of objects. }
\label{rf_gp_nan_check}
\end{figure}

To assess the robustness of the algorithms, we present a confusion matrix comparing their predictions (Figure \ref{rf_gp_nan_check}). This result reveals strong agreement between the spectral types predicted by both algorithms, with the exception of five objects. In four of these cases, the RF algorithm classifies the object class 0 (M6–M9), while GP categorizes them as class 5 (Y). For the fifth object, RF assigns it to class 1 (L0–L4), and GP identifies it again as class 5. These five objects, along with the class probabilities attributed by each algorithm, are detailed in Table \ref{tab:discrep_rf_gp}. Furthermore, the young age of these brown dwarfs in this table \citep{2001A&A...372..173B,2010A&A...515A..91M,2015ApJS..219...33G}  suggests that RF better captures the spectral features of young brown dwarfs than GP. The complete list of spectral types predicted by both algorithms for all objects in our sample is available in Table \ref{tab:predicted_types} in Appendix \ref{apend}.

\begin{table}[!ht]
\RaggedRight
\footnotesize

\caption{Prediction results for the objects showing \\ the greatest discrepancy between the outputs of both algorithms.}
\label{tab:discrep_rf_gp}
\centering
\begin{tabular}{llcccl}
\toprule
Object & Spectral  & RF  & RF  & GP & GP  \\
       &  type & predicted & prob. & predicted & prob. \\
\midrule
2MASS J0342-6817 & L4$^{(1)}$    & L0--L4 & 0.659 & Y  & 0.343 \\
GM Tau           & M6.5$^{(2)}$  & M6--M9 & 0.623 & Y  & 0.229 \\
ISO-Oph 160      & M7.75$^{(3)}$ & M6--M9 & 0.623 & Y  & 0.299 \\
ISO-Oph 164      & M8$^{(3)}$    & M6--M9 & 0.623 & Y  & 0.261 \\
ISO-Oph 176      & M7.5$^{(3)}$  & M6--M9 & 0.623 & Y  & 0.261 \\
\bottomrule
\end{tabular}

\tabletext{
(1)~\citet{2022ApJ...924...68V}; 
(2)~\citet{2021ApJ...921..182R}; 
(3)~\citet{2016A&A...593A.111T}.}

\end{table}

From this analysis, we could see that although the GP algorithm assigns spectral type Y, it does so with very low confidence. In no case did the probability exceed $35\%$, while the spectral type determined by RF, which matches the value reported in the literature, was indicated with a probability of approximately $65\%$. For this reason, although both algorithms are generally consistent, when discrepancies arise, the probabilities provided by each should be examined. The spectral type with the highest probability should then be selected as the most reliable.

\begin{table}[!ht]
\RaggedRight
\footnotesize

\caption{Predicted spectral types for isolated brown dwarfs \\ lacking previous determinations.}
\label{tab:df_new_nan}
\centering
\begin{tabular}{lcccc}
\hline \hline
Object & RF        & RF           & GP        & GP  \\
     & predicted & prob.  & predicted & prob. \\
\hline
1RXS J103137.1$-$690205 b & M6–M9 & 0.883 & M6–M9 & 0.416 \\
2M1510 Ac               & M6–M9 & 0.869 & M6–M9 & 0.660 \\
2MASS J0337$-$1758 b      & L0–L4 & 0.608 & L0–L4 & 0.396 \\
2MASS J0518$-$0153        & M6–M9 & 0.985 & M6–M9 & 0.638 \\
2MASS J1146$+$2230 b      & L0–L4 & 0.974 & L0–L4 & 0.722 \\
2MASS J1547$-$2423 b      & M6–M9 & 0.628 & L0–L4 & 0.427 \\
2MASS J1807$+$5015        & L0–L4 & 0.991 & L0–L4 & 0.701 \\
CFHT-BD-Tau 18 b        & M6–M9 & 0.623 & M6–M9 & 0.325 \\
CFHT-Pl-12 b            & M6–M9 & 0.982 & M6–M9 & 0.712 \\
EROS-MP J0032$-$4405      & L0–L4 & 0.987 & L0–L4 & 0.719 \\
IPMBD 29                & M6–M9 & 0.988 & M6–M9 & 0.703 \\
IPMBD 29 b              & M6–M9 & 0.988 & M6–M9 & 0.703 \\
ITG 25B                 & M6–M9 & 0.623 & L0–L4 & 0.201 \\
Mayrit 1701117          & M6–M9 & 0.633 & T0–T4 & 0.234 \\
PSO J228.6$-$29           & L0–L4 & 0.832 & L0–L4 & 0.606 \\
{[}SCH06{]} J0359$+$2009 b& M6–M9 & 0.997 & M6–M9 & 0.674 \\
S Ori 36                & M6–M9 & 0.918 & M6–M9 & 0.631 \\
S Ori J053854.9$-$024034  & M6–M9 & 0.782 & M6–M9 & 0.541 \\
S Ori J053855.4$-$024121  & M6–M9 & 0.629 & M6–M9 & 0.391 \\
S Ori J053918.1$-$025257  & M6–M9 & 0.892 & M6–M9 & 0.654 \\
S Ori J053922.2$-$024552  & M6–M9 & 0.939 & M6–M9 & 0.629 \\
\hline
\end{tabular}
\end{table}

Continuing with the same procedure, we applied both classification algorithms to the 21 isolated brown dwarfs for which no spectral type had been previously established, subsequently comparing the results. Table \ref{tab:df_new_nan} indicates strong agreement between the spectral types predicted by both algorithms, with the majority of objects classified as class 0 (M6–M9) and class 1 (L0–L4). We identified only three discrepancies, corresponding to 2MASS J1547$-$2423 b, ITG 25B, and Mayrit 1701117. The first two were classified as class 1 (L0–L4) by the GP algorithm and as class 0 (M6–M9) by RF, whereas Mayrit 1701117 was classified as class 3 (T0–T4) by GP and as class 0 (M6–M9) by RF. It should be noted that these brown dwarfs are young and belong to star-forming regions \citep{1999AJ....117.1471I, 2008A&A...478..667C, 2019A&A...627A.167C}. In all three cases, as shown in Table \ref{tab:df_new_nan}, the RF algorithm assigned the spectral types with higher probabilities, approximately $60\%$. This is consistent with the previous discussion that the RF classifier better accounts for the spectral features present in young brown dwarfs.  For this reason, we assign the spectral type M6-M9 to all three objects.

\section{Summary and conclusions}

In this work, we investigated the possibility of classifying brown dwarfs on the basis of photometric data using machine learning techniques. We utilized a sample of 1723 isolated brown dwarfs with known spectral types from the UltracoolSheet catalogue and analysed the relationship between various colour indices and spectral type. 

To overcome the limitations of working exclusively with colour-colour diagrams, we implemented two supervised machine learning algorithms—Random Forest and Gaussian Processes—to classify brown dwarfs based on a set of selected magnitudes and colour indices. We grouped the spectral types into six broader classes to better account for the dispersion in brown dwarf colours and the heterogeneous distribution of individuals by spectral subtype. The selected features were the absolute  magnitudes in the J, H, and W1 bands, as well as colour indices W1$-$W2, J$-$H, J$-$W1, and J$-$W2. Both algorithms achieved high classification performance, with F1-scores of 0.86 for Random Forest and 0.87 for Gaussian Processes.

The models were evaluated not only on a test set, which was separated from the original UltracoolSheet catalogue sample, but also on an independent external sample of 196 directly imaged brown dwarfs with known spectral types. Both classifiers demonstrated strong performance on this new sample, exhibiting over $80\%$ agreement.

The primary limitation of our approach is the lack of sufficient training examples for the later spectral types, especially Y dwarfs. This class imbalance increases the classification error for those types and reduces the model's generalization capability.

Among the brown dwarfs detected via direct imaging and listed in the Exoplanetary Systems Encyclopaedia, 21 lacked a previously reported spectral type. Using our trained classifiers, we successfully estimated their spectral types, and both models showed high consistency in their predictions.

Our results show that spectral classification of brown dwarfs using only photometric data is both viable and effective. This is particularly valuable for objects too faint for spectroscopic observations. By incorporating multiple colour indices simultaneously, machine learning classifiers can capture complex patterns that are not visible in traditional diagrams. 

The approach presented in this work is highly applicable to future near-infrared surveys conducted by the Roman Space Telescope \citep{2023arXiv230612363H}. These programs are expected to yield a large number of brown dwarfs that lack spectroscopic characterization, for which complementary mid-infrared data could be obtained through follow-up observations with the James Webb Space Telescope \citep{2024ApJ...976...82T}. Furthermore, although the Vera C. Rubin Observatory’s LSST is primarily an optical photometric survey, the sheer volume of data and the depth achievable with stacked images imply that a significant number of substellar objects may be discovered \citep{2022ApJS..263...23G}.

Once the current framework is validated at optical wavelengths, it can be extended to the LSST photometric system to provide initial classifications for sources missing spectra. However, applying this scheme to the survey requires addressing the inherent colour degeneracy between late-L- and T-type dwarfs and their primary contaminants: reddened main-sequence stars and high-redshift quasars ($z > 5.8$). While interstellar extinction mimics low effective temperatures, the Lyman-$\alpha$ forest break in high-$z$ QSOs can closely resemble the strong alkali metal absorption seen in the $i$ and $z$ bands of brown dwarf spectra. To alleviate this overlap in colour space, multi-band photometry must be integrated with the observatory's native proper motion and parallax measurements—distinguishing local sources from extragalactic ones—as well as time-domain variability \citep{2015A&A...574A..78S, 2021A&A...654A..79G, 2024ApJS..275...19Y}.

\section{ACKNOWLEDGEMENTS}
The authors would like to thank the anonymous reviewer who performed a careful and detailed reading of the manuscript; their constructive comments and suggestions significantly improved the scientific discussion and overall presentation of this article.

----------------
\bibliographystyle{rmaa}
\bibliography{referencias}

\begin{appendices}

\section{Appendix: Spectral types predicted by the RF and GP algorithms}
\label{apend}

List of 196 directly imaged brown dwarfs with previous spectral types from the Encyclopaedia of Exoplanetary Systems. These targets possess all the features required for our RF and GP classification algorithms. Column 1: Name; Column 2: RF predicted class; Column 3: RF probability; Column 4: GP predicted class; Column 5: GP probability.

\begin{table*}
\centering

\caption{Spectral type predictions from RF and GP algorithms}
\label{tab:predicted_types} 
\begin{tabular}{lcccc}
\hline
\hline
Object & RF predicted     & RF probability     & GP predicted    & GP probability \\
\hline

2M1510 B & M6-M9 & 0.87 & M6-M9 & 0.66 \\
2MASS J0001+1535 & L0-L4 & 0.79 & L0-L4 & 0.58 \\
2MASS J0004-6410 & L0-L4 & 0.79 & L0-L4 & 0.69 \\
2MASS J0031+5749 & T0-T4 & 0.46 & L0-L4 & 0.40 \\
2MASS J00413538-5621127 A & M6-M9 & 1.00 & M6-M9 & 0.71 \\
2MASS J00413538-5621127 B & M6-M9 & 1.00 & M6-M9 & 0.71 \\
2MASS J0117-3403 & L0-L4 & 0.98 & L0-L4 & 0.71 \\
2MASS J0153-6744 & L0-L4 & 0.64 & L0-L4 & 0.56 \\
2MASS J0207+0000 & T0-T4 & 0.82 & T0-T4 & 0.57 \\
2MASS J0210-3015 & L0-L4 & 0.97 & L0-L4 & 0.72 \\
2MASS J0223-5815 & L0-L4 & 0.97 & L0-L4 & 0.71 \\
2MASS J0226-5327 & L0-L4 & 0.96 & L0-L4 & 0.71 \\
2MASS J0234-6442 & L0-L4 & 0.98 & L0-L4 & 0.72 \\
2MASS J0241-0326 & L0-L4 & 0.92 & L0-L4 & 0.72 \\
2MASS J0253+3206 & M6-M9 & 0.95 & M6-M9 & 0.71 \\
2MASS J0310+1648 & L5-L9 & 0.82 & L5-L9 & 0.50 \\
2MASS J0310+1648 b & L5-L9 & 0.82 & L5-L9 & 0.50 \\
2MASS J0325+0425 & T5-T9 & 0.91 & T5-T9 & 0.76 \\
2MASS J0326-2102 & L0-L4 & 0.52 & L5-L9 & 0.42 \\
2MASS J0328+2302 & L5-L9 & 0.98 & L5-L9 & 0.65 \\
2MASS J0337-1758 & L0-L4 & 0.61 & L0-L4 & 0.40 \\
2MASS J0342-6817 & L0-L4 & 0.66 & Y & 0.34 \\
2MASS J0407+1546 & L0-L4 & 0.88 & L0-L4 & 0.65 \\
2MASS J0414+2811 & M6-M9 & 0.62 & M6-M9 & 0.32 \\
2MASS J0418+2131 & L5-L9 & 0.63 & L5-L9 & 0.42 \\
2MASS J0423+2801 & M6-M9 & 0.89 & M6-M9 & 0.58 \\
2MASS J0439+2336 & M6-M9 & 0.91 & M6-M9 & 0.56 \\
2MASS J0439-2353 & L5-L9 & 0.87 & L5-L9 & 0.51 \\
2MASS J0440+2358 & M6-M9 & 0.94 & M6-M9 & 0.60 \\
2MASS J0441+2534 & M6-M9 & 0.63 & M6-M9 & 0.45 \\
2MASS J0444+2512 & M6-M9 & 0.63 & M6-M9 & 0.48 \\
2MASS J0447-1216 & T0-T4 & 0.87 & T0-T4 & 0.62 \\
2MASS J0501-0010 & L0-L4 & 0.45 & L5-L9 & 0.43 \\
2MASS J0502+1442 & L0-L4 & 0.83 & L0-L4 & 0.42 \\
2MASS J0506+5236 & T0-T4 & 0.96 & T0-T4 & 0.64 \\
2MASS J0518-2756 & L0-L4 & 0.88 & L0-L4 & 0.72 \\
2MASS J0523-1403 & L0-L4 & 0.99 & L0-L4 & 0.71 \\
2MASS J0533-0156 & M6-M9 & 0.88 & M6-M9 & 0.62 \\
2MASS J0537-0155 & M6-M9 & 0.97 & M6-M9 & 0.66 \\
2MASS J0608-2753 & L0-L4 & 0.52 & L0-L4 & 0.54 \\
2MASS J0614+3950 & L5-L9 & 0.66 & T0-T4 & 0.40 \\
2MASS J0850+1057 b & L5-L9 & 0.91 & L5-L9 & 0.63 \\
2MASS J0856-1342 & M6-M9 & 0.63 & M6-M9 & 0.60 \\
2MASS J0856+3746 & L0-L4 & 0.72 & L0-L4 & 0.46 \\
2MASS J0949-1545 & T0-T4 & 0.98 & T0-T4 & 0.68 \\
2MASS J0951-8023 & L0-L4 & 0.82 & L0-L4 & 0.58 \\
2MASS J11011926-7732383 A & M6-M9 & 0.93 & M6-M9 & 0.60 \\
2MASS J11011926-7732383 B & M6-M9 & 0.93 & M6-M9 & 0.60 \\
2MASS J1115+1937 & L0-L4 & 0.56 & L0-L4 & 0.57 \\
2MASS J1146+2230 & L0-L4 & 0.97 & L0-L4 & 0.72 \\
2MASS J1200-7845 & M6-M9 & 0.89 & M6-M9 & 0.60 \\
2MASS J1247-3816 & M6-M9 & 0.60 & M6-M9 & 0.46 \\

\hline
\end{tabular}
\end{table*}

\begin{table*}
\centering
\captionsetup{list=no}
\caption*{Table 4 — Continued}

\begin{tabular}{lcccc}
\hline
\hline
Object & RF predicted     & RF probability     & GP predicted    & GP probability \\
\hline

2MASS J1341-3052 & L0-L4 & 0.99 & L0-L4 & 0.71 \\
2MASS J1341-3052 b & L0-L4 & 0.99 & L0-L4 & 0.71 \\
2MASS J1411-2119 & M6-M9 & 0.98 & M6-M9 & 0.70 \\
2MASS J1448+1031 & L0-L4 & 0.85 & L0-L4 & 0.59 \\
2MASS J1516+0259 & M6-M9 & 0.54 & L0-L4 & 0.39 \\
2MASS J1516+0259 b & M6-M9 & 0.54 & L0-L4 & 0.39 \\
2MASS J1516+3053 & T0-T4 & 0.63 & L5-L9 & 0.50 \\
2MASS J1516+3053 b & T0-T4 & 0.63 & L5-L9 & 0.50 \\
2MASS J1546-3325 & T5-T9 & 1.00 & T5-T9 & 0.72 \\
2MASS J1547-2423 & M6-M9 & 0.63 & L0-L4 & 0.43 \\
2MASS J1551+0941 & L0-L4 & 0.62 & L0-L4 & 0.62 \\
2MASS J1552+2948 & L0-L4 & 0.98 & L0-L4 & 0.72 \\
2MASS J1617-1858 & M6-M9 & 0.54 & L0-L4 & 0.39 \\
2MASS J1711+2232 & L5-L9 & 0.98 & L5-L9 & 0.69 \\
2MASS J1711+2232 b & L5-L9 & 0.98 & L5-L9 & 0.69 \\
2MASS J1726+1538 & L0-L4 & 0.60 & L0-L4 & 0.65 \\
2MASS J1750+1759 & T0-T4 & 0.96 & T0-T4 & 0.64 \\
2MASS J1841+3117 & L0-L4 & 0.58 & L0-L4 & 0.55 \\
2MASS J2033-5635 & L0-L4 & 0.92 & L0-L4 & 0.67 \\
2MASS J2057-0252 & L0-L4 & 0.99 & L0-L4 & 0.72 \\
2MASS J2202-5605 & M6-M9 & 0.86 & M6-M9 & 0.65 \\
2MASS J2202-5605 b & M6-M9 & 0.86 & M6-M9 & 0.65 \\
2MASS J2206-4217 & L0-L4 & 0.73 & L0-L4 & 0.47 \\
2MASS J2322-6151 & L0-L4 & 0.81 & L0-L4 & 0.70 \\
2MASS J2322-6151 b & L0-L4 & 0.81 & L0-L4 & 0.70 \\
2MASS J2343-3646 & L5-L9 & 0.68 & L5-L9 & 0.57 \\
BRB 4 & M6-M9 & 0.99 & M6-M9 & 0.71 \\
CFHT-BD-Tau 1 & M6-M9 & 0.83 & M6-M9 & 0.49 \\
CFHT-BD-Tau 18 & M6-M9 & 0.62 & M6-M9 & 0.33 \\
CFHT-BD-Tau 2 & M6-M9 & 0.94 & M6-M9 & 0.60 \\
CFHT-BD-Tau 3 & M6-M9 & 0.90 & M6-M9 & 0.63 \\
CFHT-BD-Tau 4 & M6-M9 & 0.63 & M6-M9 & 0.39 \\
CFHT-BD-Tau 6 & M6-M9 & 0.84 & M6-M9 & 0.58 \\
CFHT-PL-11 & M6-M9 & 0.83 & M6-M9 & 0.71 \\
CFHT-Pl-12 & M6-M9 & 0.98 & M6-M9 & 0.71 \\
CFHT-PL-15 & M6-M9 & 0.55 & M6-M9 & 0.53 \\
CFHT-PL-5 & M6-M9 & 1.00 & M6-M9 & 0.71 \\
CFHT-PL-6 & M6-M9 & 0.99 & M6-M9 & 0.69 \\
CFHTWIR-Oph 58 & L0-L4 & 0.59 & L0-L4 & 0.35 \\
Cha Ha 1 & M6-M9 & 0.85 & M6-M9 & 0.60 \\
Cha Ha 11 & M6-M9 & 0.85 & M6-M9 & 0.60 \\
Cha Ha 12 & M6-M9 & 0.97 & M6-M9 & 0.60 \\
DENIS 1538-1038 & M6-M9 & 0.87 & M6-M9 & 0.70 \\
DENIS 19 & L0-L4 & 0.94 & L0-L4 & 0.67 \\
DENIS 19 b & L0-L4 & 0.94 & L0-L4 & 0.67 \\
DENIS J0225-5837 & M6-M9 & 0.81 & M6-M9 & 0.60 \\
DENIS J0255-4700 & L5-L9 & 0.91 & L5-L9 & 0.63 \\
DENIS J185950.9-370632 & M6-M9 & 0.88 & M6-M9 & 0.61 \\
DENIS J2000-7523 & M6-M9 & 0.78 & M6-M9 & 0.59 \\
FLMN J0541506-0158041 & M6-M9 & 0.86 & M6-M9 & 0.49 \\
GKH94-41 & L0-L4 & 0.45 & L5-L9 & 0.19 \\
GM Tau & M6-M9 & 0.62 & Y & 0.23 \\

\hline
\end{tabular}
\end{table*}

\begin{table*}
\centering
\captionsetup{list=no}
\caption*{Table 4 — Continued}
\begin{tabular}{lcccc}
\hline
\hline
Object & RF predicted     & RF probability     & GP predicted    & GP probability \\
\hline

GO Cet & L5-L9 & 0.91 & L5-L9 & 0.60 \\
GY 11 & M6-M9 & 0.92 & M6-M9 & 0.55 \\
GY 141 & M6-M9 & 0.92 & M6-M9 & 0.55 \\
GY 202 & L0-L4 & 0.60 & L5-L9 & 0.23 \\
GY 264 & M6-M9 & 0.80 & M6-M9 & 0.57 \\
GY 320 & M6-M9 & 0.92 & M6-M9 & 0.57 \\
HD 130948 C & M6-M9 & 0.80 & M6-M9 & 0.40 \\
HHJ 3 & M6-M9 & 1.00 & M6-M9 & 0.71 \\
ISO-Oph 138 & L0-L4 & 0.59 & L5-L9 & 0.22 \\
ISO-Oph 160 & M6-M9 & 0.62 & Y & 0.30 \\
ISO-Oph 164 & M6-M9 & 0.62 & Y & 0.26 \\
ISO-Oph 176 & M6-M9 & 0.62 & Y & 0.26 \\
KG2001-102 & M6-M9 & 0.83 & M6-M9 & 0.62 \\
KPNO-Tau 1 & M6-M9 & 0.81 & M6-M9 & 0.64 \\
KPNO-Tau 2 & M6-M9 & 0.94 & M6-M9 & 0.66 \\
KPNO-Tau 4 & L0-L4 & 0.52 & L0-L4 & 0.53 \\
KPNO-Tau 5 & M6-M9 & 0.88 & M6-M9 & 0.61 \\
KPNO-Tau 6 & L0-L4 & 0.52 & L0-L4 & 0.53 \\
KPNO-Tau 7 & M6-M9 & 0.78 & M6-M9 & 0.58 \\
KPNO-Tau 9 & M6-M9 & 0.58 & M6-M9 & 0.64 \\
LH 0419+15 & M6-M9 & 0.93 & M6-M9 & 0.62 \\
LS-RCrA 1 & M6-M9 & 0.82 & M6-M9 & 0.59 \\
MHO-Tau-4 & L0-L4 & 0.65 & M6-M9 & 0.40 \\
MHO-Tau-5 & M6-M9 & 0.80 & M6-M9 & 0.70 \\
Par-Lup3-1 & M6-M9 & 0.93 & M6-M9 & 0.60 \\
PSO J229.2-26 & L0-L4 & 0.60 & L0-L4 & 0.49 \\
PSO J231.7-26 & M6-M9 & 0.57 & M6-M9 & 0.40 \\
PSO J231.8-29 & L0-L4 & 0.50 & L0-L4 & 0.41 \\
PSO J237.1-23 & M6-M9 & 0.82 & M6-M9 & 0.49 \\
PSO J239.7-23 & M6-M9 & 0.58 & L0-L4 & 0.51 \\
PSO J316.1-09 & L0-L4 & 0.99 & L0-L4 & 0.71 \\
Roque 12 & M6-M9 & 0.80 & M6-M9 & 0.57 \\
Roque 13 & M6-M9 & 0.85 & M6-M9 & 0.61 \\
SCH06-J0359+2009 & M6-M9 & 1.00 & M6-M9 & 0.67 \\
SDSS J0926+5847 & T0-T4 & 0.63 & T0-T4 & 0.56 \\
SDSS J0926+5847 b & T0-T4 & 0.63 & T0-T4 & 0.56 \\
SDSS J1021-0304 & T0-T4 & 0.97 & T0-T4 & 0.74 \\
SDSS J1021-0304 b & T0-T4 & 0.97 & T0-T4 & 0.74 \\
SDSS J1039+3256 b & T0-T4 & 0.69 & T0-T4 & 0.68 \\
SDSS J1052+4422 & L5-L9 & 0.54 & T0-T4 & 0.57 \\
SDSS J1052+4422 b & L5-L9 & 0.54 & T0-T4 & 0.57 \\
SDSS J1416+1348 & L5-L9 & 0.98 & L5-L9 & 0.70 \\
SDSS J1416+1348 b & L5-L9 & 0.98 & L5-L9 & 0.70 \\
SDSS J1534+1615 & T0-T4 & 0.98 & T0-T4 & 0.75 \\
SDSS J1534+1615 b & T0-T4 & 0.98 & T0-T4 & 0.75 \\
SDSS J2249+0044 b & L0-L4 & 0.57 & L0-L4 & 0.48 \\
SI2M-37 & M6-M9 & 0.83 & M6-M9 & 0.54 \\
SIMP J1501-0135 & L0-L4 & 0.75 & L0-L4 & 0.65 \\
SIMP J1501-0135 b & L0-L4 & 0.75 & L0-L4 & 0.65 \\
SIMP J1619+0313 & T0-T4 & 0.94 & T0-T4 & 0.76 \\
SIMP J1619+0313 b & T0-T4 & 0.94 & T0-T4 & 0.76 \\
SIMP J2154-1055 & L0-L4 & 0.33 & L0-L4 & 0.56 \\
\hline
\end{tabular}
\end{table*}

\begin{table*}
\centering
\captionsetup{list=no}
\caption*{Table 4 — Continued}
\begin{tabular}{lcccc}
\hline
\hline
Object & RF predicted     & RF probability     & GP predicted    & GP probability \\
\hline

SIPS J0058-0651 & L0-L4 & 0.87 & L0-L4 & 0.68 \\
SIPS J0847-1532 & L0-L4 & 0.99 & L0-L4 & 0.72 \\
SONYC-Lup3-7 & M6-M9 & 0.82 & M6-M9 & 0.58 \\
S Ori 20 & M6-M9 & 0.98 & M6-M9 & 0.62 \\
S Ori 25 & M6-M9 & 0.99 & M6-M9 & 0.60 \\
S Ori 27 & M6-M9 & 0.98 & M6-M9 & 0.61 \\
S Ori 28 & M6-M9 & 0.98 & M6-M9 & 0.64 \\
S Ori 31 & M6-M9 & 1.00 & M6-M9 & 0.64 \\
S Ori 32 & M6-M9 & 0.99 & M6-M9 & 0.64 \\
S Ori 42 & M6-M9 & 0.80 & M6-M9 & 0.60 \\
S Ori 45 & M6-M9 & 0.84 & M6-M9 & 0.65 \\
S Ori 47 & M6-M9 & 0.56 & L0-L4 & 0.39 \\
S Ori J053825.4-024241 & M6-M9 & 0.78 & M6-M9 & 0.58 \\
S Ori J053826.1-024041 & M6-M9 & 1.00 & M6-M9 & 0.63 \\
S Ori J053829.0-024847 & M6-M9 & 0.91 & M6-M9 & 0.60 \\
S Ori J053954.3-023719 & M6-M9 & 0.98 & M6-M9 & 0.61 \\
S Ori J054004.5-023642 & M6-M9 & 0.89 & M6-M9 & 0.60 \\
SSTc2d J163134.1 & L0-L4 & 0.41 & L5-L9 & 0.21 \\
Teide 2 & M6-M9 & 0.96 & M6-M9 & 0.71 \\
TOI-456 & M6-M9 & 0.96 & M6-M9 & 0.71 \\
TVLM 831-154910 & M6-M9 & 0.99 & M6-M9 & 0.71 \\
TWA 26 & M6-M9 & 0.92 & M6-M9 & 0.68 \\
TWA 28 & M6-M9 & 0.82 & M6-M9 & 0.60 \\
UGPS J0521+3640 & L0-L4 & 0.71 & L0-L4 & 0.65 \\
USco 1606-2335 & M6-M9 & 0.77 & M6-M9 & 0.54 \\
USco 1607-2211 & M6-M9 & 0.74 & M6-M9 & 0.63 \\
USco 1607-2242 & L0-L4 & 0.89 & L0-L4 & 0.63 \\
USco 1608-2232 & M6-M9 & 0.62 & L0-L4 & 0.43 \\
USco 1608-2315 & L0-L4 & 0.62 & M6-M9 & 0.51 \\
USco 1610-2239 & M6-M9 & 0.62 & M6-M9 & 0.59 \\
UScoCTIO 100 & M6-M9 & 0.98 & M6-M9 & 0.62 \\
UScoCTIO 109 & L0-L4 & 0.59 & L0-L4 & 0.35 \\
UScoCTIO 128 & M6-M9 & 0.83 & M6-M9 & 0.59 \\
UScoCTIO 130 & M6-M9 & 0.94 & M6-M9 & 0.68 \\
UScoCTIO 131 & M6-M9 & 0.97 & M6-M9 & 0.69 \\
UScoCTIO 137 & M6-M9 & 0.78 & M6-M9 & 0.69 \\
USco J1554-2135 & M6-M9 & 0.59 & M6-M9 & 0.50 \\
WISE J1506+7027 & T0-T4 & 0.31 & T0-T4 & 0.23 \\
WISE J1810-1010 & T0-T4 & 0.62 & T0-T4 & 0.40 \\
WISE J1925+0700 & L5-L9 & 0.94 & L5-L9 & 0.70 \\
\hline
\end{tabular}
\end{table*}

\end{appendices}

\end{document}